\documentstyle[preprint,eqsecnum,aps]{revtex}
\begin{document}
\draft

%======================================%
%<<<<<<<<<<<< TITLE PAGE >>>>>>>>>>>>>>%
%======================================%
\preprint{YITP-97-59, gr-qc/9711058}
\title{
Excitation of a Kaluza-Klein mode by parametric resonance
}
\author{Shinji Mukohyama}
\address{
Yukawa Institute for Theoretical Physics, 
Kyoto University \\
Kyoto 606-01, Japan
}
\date{\today}

\maketitle

%======================================%
%<<<<<<<<<<<<< ABSTRACT >>>>>>>>>>>>>>>% 
%======================================%

\begin{abstract} 

In this paper we investigate a parametric resonance phenomenon of a
Kaluza-Klein mode in a $D$-dimensional generalized Kaluza-Klein
theory. As the origin of the parametric resonance we consider a small
oscillation of a scale of the compactification around a today's value
of it. To make our arguments definite and for simplicity we consider
two classes of models of the compactification: those by $S_{d}$
($d=D-4$) and those by $S_{d_{1}}\times S_{d_{2}}$ ($d_1\ge d_2$,
$d_{1}+d_{2}=D-4$). For these models we show that parametric resonance
can occur for the Kaluza-Klein mode. After that, we give formulas of
a creation rate and a number of created quanta of the Kaluza-Klein
mode due to the parametric resonance, taking into account the first
and the second resonance band. By using the formulas we calculate those 
quantities for each model of the compactification. Finally we give
conditions for the parametric resonance to be efficient and discuss 
cosmological implications. 

\end{abstract}


\newpage
%\vskip1cm

%======================================%
%<<<<<<<<<<<< SECTION I  >>>>>>>>>>>>>>%
%======================================%

\section{Introduction}\label{sec:intro}

Many unified theories require spacetime dimension higher than $4$. We 
call them generalized Kaluza-Klein theories. For example the superstring 
theory predicts spacetime dimension $D=10$~\cite{GSW}, the M-theory 
$D=11$~\cite{M-theory}. On the contrary we can see only a
$4$-dimensional part of the spacetime by experiments. Hence  
the extra $(D-4)$-dimensional part is thought to be compactified so that 
it cannot be seen by us. One may be able to give a $4$-dimensional theory 
describing our $4$-dimensional universe from a generalized
Kaluza-Klein theory by the compactification of the extra
dimensions. However, there are a plenty 
number of ways of compactification and there are the same number of 
the corresponding $4$-dimensional theories for a $D$-dimensional 
theory. Thus we need  criterion to judge which compactification must
be selected. For example, in the case of the superstring theory it is 
well known that a `good' compactification must be realized by a
Calabi-Yau manifold in order for the $N=1$ supersymmetry to remain in
the corresponding $4$-dimensional theory~\cite{GSW}. In this paper we 
give a suggestive argument toward a cosmological criterion for
compactification.

Kolb and Slansky~\cite{Kolb&Slansky} examined a model of the 
compactification by $S_{1}$ in the $5$-dimensional theory with a
metric field described by an Einstein-Hilbert action and a massless
scalar field described by a Klein-Gordon action. They investigated 
a cosmological evolution of energy density of a mode of the scalar
field with a non-zero momentum in the direction of the
compactified extra space $S_{1}$ assuming that there is no entropy
production. The mode is usually called a Kaluza-Klein mode and the 
quantum of the mode is called a pyrgon. They showed that if a scale of
the compactification is of order of the $4$-dimensional Planck length
and energy density of pyrgons is comparable with energy density of
radiation at an early epoch of the universe, then at the recombination
epoch the former energy density terribly exceeds the critical density
of the universe. The physical reason of their result is essentially
the momentum conservation in the direction of the compactification: a
pyrgon with a positive momentum in the direction cannot decay without
meeting with another pyrgon with a negative momentum. For this reason 
energy density $\rho_{KK}$ of the Kaluza-Klein mode decreases as 
$a^{-3}$, where $a$ is a scale factor of the universe, while energy 
density $\rho_{rad}$ of radiation decreases as $a^{-4}$. As a result 
the ratio $\rho_{KK}/\rho_{rad}$ evolves as $a$. Since the momentum
conservation in the direction of the compactification does always
hold, their result is expected to hold in a wider class of models of
the compactification. Thus, if entropy production is negligible,
energy density of pyrgons must not be comparable with energy density
of radiation at any early epoch of the universe in a wider class of
models of the compactification. On the other hand, if pyrgons are
suitably created at an early epoch and they are diluted by entropy
production after their creation (eg. reheating after inflation) then
the pyrgons may become a dark matter at today. Anyway, we have to
analyze whether pyrgons can be created in early universe or not in
order to obtain a knowledge of the compactification from a
cosmological point of view. In this paper, motivated by these
considerations, we investigate a catastrophic creation of quanta of a
Kaluza-Klein mode in a $D$-dimensional generalized Kaluza-Klein
theory.

As the origin of the catastrophic creation of quanta of the
Kaluza-Klein mode we consider a small oscillation of a scale of the
compactification at an early epoch of the universe. In general it is
expected that frequency of the oscillation is of order of inverse of
the scale of the compactification and mass of the Kaluza-Klein mode is
of the same order. Hence the mass of the Kaluza-Klein mode oscillates
with a frequency of order of the mass itself. Such an oscillation of
the mass causes a catastrophic creation of quanta when frequency and
amplitude of the oscillation are in a band in their parameter
space. This sort of catastrophic creation of quanta is called a
parametric resonance phenomenon~\cite{Landau}. In this paper we apply
a theory of parametric resonance~\footnote{
The theory of parametric resonance was applied to reheating (or
preheating) process after inflation by many
authors~\cite{Reheating,STB}.
}
to a Kaluza-Klein mode and judge whether the 
catastrophic creation occurs efficiently or not.

In Sec. \ref{sec:action} we calculate $4$-dimensional actions for 
two classes of simple models of the compactification. In
Sec. \ref{sec:resonance} a Kaluza-Klein mode is investigated quantum
mechanically on a classical background of a $4$-dimensional Friedmann
universe with a small oscillation of the scale of the
compactification. We show that parametric resonance of the
Kaluza-Klein mode can occur. After that we intend to judge whether the
resonance is efficient to produce sufficient quanta of the
Kaluza-Klein mode. Sec. \ref{sec:discussion} is devoted to summarize
this paper.

%======================================%
%<<<<<<<<<<<< SECTION II >>>>>>>>>>>>>>%
%======================================%

\section{$4$-dimensional action}\label{sec:action}

Our starting point in this paper is a $D$-dimensional Einstein-Hilbert 
action with a cosmological constant $\bar{\Lambda}$:
%============< EQUATION >==============%
%
\begin{equation}
 I_{EH} = \frac{1}{2\bar{\kappa}^2}
		\int d^{D}x\sqrt{-\bar{g}}(\bar{R}-2\bar{\Lambda}),
\end{equation}
%======================================%
where $\bar{\kappa}$ is a positive constant, $\bar{g}$ and $\bar{R}$
are a determinant and a Ricci scalar of a $D$-dimensional metric tensor
$\bar{g}_{MN}$~\footnote{
Hereafter we take the sign convention $(-++\cdots +)$ of the metric 
tensor.
}. 
In addition to the gravitational field described by the Einstein-Hilbert 
action, we consider other massless fields, too. For the present, we
include a real massless scalar field $\bar{\phi}$ described by a 
$D$-dimensional Klein-Gordon action~\footnote{
Here we mention that the low energy action of the superstring theory
and the M-theory can be reduced to a form which includes the
Einstein-Hilbert action and the massless Klein-Gordon action.
}:
%============< EQUATION >==============%
%
\begin{equation}
 I_{KG} = -\frac{1}{2}\int d^{D}x\sqrt{-\bar{g}}\bar{g}^{MN}
 	\partial_{M}\bar{\phi}\partial_{N}\bar{\phi}.
\end{equation}
%======================================%
As stated in the previous section, some candidates of the unified theory 
require $D>4$ while the spacetime dimension we can see by experiments is 
$4$. Hence we have to compactify the extra $(D-4)$-dimensional space so 
that it can not be seen by experiments. In this paper to make arguments 
definite we consider two classes of simple models of the 
compactification. Unfortunately, there are a great number of other models 
of the compactification and there is no conclusive first principle to
determine which model should be taken. However, qualitative properties 
shown in this paper for simple models are expected to be common in a 
wider class of models of the compactification.

\subsection{Compactification by $S_{d}$}

First we consider a compactification by a $d$-dimensional sphere 
$S_{d}$ $(d=D-4)$. In this case the metric tensor $\bar{g}_{MN}$ can 
be written as
%============< EQUATION >==============%
%
\begin{equation}
 \bar{g}_{MN}dx^Mdx^N = \hat{g}_{\mu\nu}dx^{\mu}dx^{\nu}
	+ b^2\Omega^{(d)}_{ij}dx^idx^j,
\end{equation}
%======================================%
where $\hat{g}_{\mu\nu}$ is a $4$-dimensional metric depending only on 
the $4$-dimensional coordinates $x^{\mu}$ $(\mu=0,1,2,3)$, $b$ is a scale 
of the compactification (a radius of the $d$-dimensional sphere) 
depending only on the $4$-dimensional coordinates, and 
$\Omega^{(d)}_{ij}dx^idx^j$ is a line element of a unit $d$-sphere. The 
scalar field $\bar{\phi}$ is expanded as follows: 
%============< EQUATION >==============%
%
\begin{equation}
 \bar{\phi} = b_{0}^{-d/2}
	\sum_{l,m}\phi_{lm}(x^{\mu})Y^{(d)}_{lm}(x^i),
\end{equation}
%======================================%
where the constant $b_{0}$ is a today's value of $b$. In the 
expression, $Y^{(d)}_{lm}$ ($l=1,2,\cdots$; and $m$ denotes a set of $d-1$ 
numbers that are need in order for a set of all $Y^{(d)}_{lm}$ to be a 
complete set of $L^2$ functions on the $d$-sphere) are real harmonics 
on the $d$-sphere satisfying 
%============< EQUATION >==============%
%
\begin{eqnarray}
 \frac{1}{\sqrt{\Omega^{(d)}}}\partial_{i}\left(
 	\sqrt{\Omega^{(d)}}\Omega^{(d)ij}\partial_{j}Y^{(d)}_{lm}\right)
	 + l(l+d-1)Y^{(d)}_{lm} 
 & = & 0 ,\\
 \int d^dx\sqrt{\Omega^{(d)}}Y^{(d)}_{lm}Y^{(d)}_{l'm'} 
 & = & \delta_{ll'}\delta_{mm'},
\end{eqnarray}
%======================================%
and $\phi_{lm}(x^{\mu})$ is a real function depending only on 
the $4$-dimensional coordinates $x^{\mu}$. The Einstein-Hilbert 
action and the Klein-Gordon action in this case are
%============< EQUATION >==============%
%
\begin{eqnarray}
 I_{EH} & = & \frac{1}{2\kappa^2}\int d^4x\sqrt{-\hat{g}}
	\left(\frac{b}{b_{0}}\right)^d\left[ \hat{R} + 
	d(d-1)\hat{g}^{\mu\nu}(\partial_{\mu}\ln b)(\partial_{\nu}\ln b)
	+ d(d-1)b^{-2} -2\bar{\Lambda}\right] ,\nonumber\\
 I_{KG} & = & -\frac{1}{2}\sum_{l,m}\int d^4x\sqrt{-\hat{g}}
 	\left(\frac{b}{b_{0}}\right)^d\left[
 	\hat{g}^{\mu\nu}\partial_{\mu}\phi_{lm}
 	\partial_{\nu}\phi_{lm} 
 	+ \frac{l(l+d-1)}{b^2}\phi_{lm}^2
 	\right],
\end{eqnarray}
%======================================%
where $\kappa$ is a positive constant defined by 
$\kappa^2=\bar{\kappa}^2/(2^Db_{0}^D\pi)$ and $\hat{R}$ is a Ricci scalar 
of the $4$-dimensional metric tensor $\hat{g}_{\mu\nu}$. Since we shall 
consider the case when the scale $b$ of the compactification is time dependent, 
the factor $(b/b_{0})^D$ before the Ricci scalar $\hat{R}$ and before the 
kinetic term
$\hat{g}^{\mu\nu}\partial_{\mu}\phi_{lm}\partial_{\nu}\phi_{lm}$ 
of the $\phi_{lm}$ field is not a constant. Hence those expressions of
the Einstein-Hilbert action and the Klein-Gordon action are quite 
different from the usual form of the $4$-dimensional Einstein-Hilbert
action and the $4$-dimensional Klein-Gordon action.

In this paper we shall examine excitations of the $\phi_{lm}$ fields
due to a time dependence of the compactification scale $b$. For this 
purpose we have to analyze a quantum dynamics of the 
$\phi_{lm}$ fields. However, as seen in the previous paragraph, the 
action of the $\phi_{lm}$ fields looks quite different from the 
usual $4$-dimensional Klein-Gordon action and is not a useful form for
the analysis of the quantum dynamics. Thus we perform the following
conformal transformation in order to obtain a useful expression. 
%============< EQUATION >==============%
%
\begin{equation}
 \hat{g}_{\mu\nu} = \left(\frac{b}{b_{0}}\right)^{-d}g_{\mu\nu}.
 \label{eqn:conformal-tr}
\end{equation}
%======================================%
By this conformal transformation we obtain
%============< EQUATION >==============%
%
\begin{eqnarray}
 I_{EH} & = & \int d^4x\sqrt{-g}\left[\frac{1}{2\kappa^2}R
	-\frac{1}{2}g^{\mu\nu}\partial_{\mu}\sigma\partial_{\nu}\sigma 
	- U_{0}(\sigma)\right] ,\nonumber\\
 I_{KG} & = & -\frac{1}{2}\sum_{l,m}\int d^4x\sqrt{-g}\left[
 	g^{\mu\nu}\partial_{\mu}\phi_{lm}\partial_{\nu}\phi_{lm} 
 	+ M_{l}^2(\sigma)\phi_{lm}^2\right],\label{eqn:EH-KG}
\end{eqnarray}
%======================================%
where $R$ is a Ricci scalar of the $4$-dimensional metric tensor 
$g_{\mu\nu}$, the field $\sigma$ is defined by 
%============< EQUATION >==============%
%
\begin{eqnarray}
 \sigma & = & \sigma_{0}\ln\left(\frac{b}{b_{0}}\right)\ \ 
		,\nonumber\\
 \sigma_{0} & = & \sqrt{\frac{d(d+2)}{2\kappa^2}},
\end{eqnarray}
%======================================%
the mass $M_{l}(\sigma)$ of the $4$-dimensional scalar field $\phi_{lm}$ 
is given by  
%============< EQUATION >==============%
%
\begin{equation}
 M_{l}^2(\sigma ) = \frac{l(l+d-1)}{b_{0}^2}
 	e^{-(d+2)\sigma /\sigma_{0}},\label{eqn:mass}
\end{equation}
%======================================%
and the potential $U_{0}(\sigma)$ of $\sigma$ is 
%============< EQUATION >==============%
%
\begin{equation}
 U_{0}(\sigma) = \frac{\bar{\Lambda}}{\kappa^2}e^{-d\sigma /\sigma_{0}}
 	-\frac{d(d-1)}{2\kappa^2b_{0}^2}e^{-(d+2)\sigma /\sigma_{0}}.
\end{equation}
%======================================%
Here we mention that a mode with non-zero $l$ has a mass of order
$1/b_0$ and is called a Kaluza-Klein mode.

The expression (\ref{eqn:EH-KG}) can be understood as the 
$4$-dimensional Einstein-Hilbert action plus the $4$-dimensional
Klein-Gordon action with a potential. However, it can be easily
confirmed that $U_{0}(\sigma)$ has no local minimum. It means that
there is no stable compactification by $S_{d}$ if the model is not
modified any more. Therefore we have to take into account some
mechanism in order to stabilize the potential of the $\sigma$
field. In this paper as the mechanism we consider a $1$-loop effective
action (an action for so-called Casimir effects) contributed by all
$4$-dimensional matter fields. The $1$-loop effective action is of the 
following form \cite{Candelas&Weinberg}: 
%============< EQUATION >==============%
%
\begin{equation}
 I_{1loop} = -\int d^4x\sqrt{-g}V_{1loop}(\sigma),
 	\label{eqn:1loop-I-V}
\end{equation}
%======================================%
where $V_{1loop}$ is an effective potential, which in turn is a sum over 
all contributing fields~\footnote{
Strictly speaking, we have to restrict the dimension $D$ to be odd 
since otherwise the conformal anomaly arises and the arguments become 
complicated. 
}.
%============< EQUATION >==============%
%
\begin{equation}
 V_{1loop}(\sigma) = \sum_{i}c_{i}V(M_{i}),\label{eqn:1loop-V}
\end{equation}
%======================================%
where $i$ distinguishes $4$-dimensional fields contributing to the 
$1$-loop effective action, $M_{i}$ is a mass of the field specified by 
$i$, and $c_{i}$ is a numerical factor depending on what spin the field 
$i$ has~\footnote{
For example, $c_{i}=1$ for spin $0$, $c_{i}=-4$ for spin
$1/2$~\cite{Candelas&Weinberg}. 
}.
In the expression the function $V(M)$ is given by
%============< EQUATION >==============%
%
\begin{equation}
 V(M) = -\frac{1}{2}\lim_{n\to 4}(4\pi)^{-n/2}\Gamma (-n/2) M^n
		.\label{eqn:VM}
\end{equation}
%======================================%
Since squared masses of $4$-dimensional fields deduced from the 
massless $D$-dimensional theory is proportional to 
$e^{-(d+2)\sigma /\sigma_{0}}$ as the squared mass (\ref{eqn:mass}) of 
the $\phi_{lm}$ fields, the $1$-loop effective action is of the following 
form in general. 
%============< EQUATION >==============%
%
\begin{equation}
 V_{1loop}(\sigma) = Ae^{-2(d+2)\sigma /\sigma_{0}}
 	,\label{eqn:V1loop}
\end{equation}
%======================================%
where $A$ is a constant. We could calculate the constant $A$ in principle 
if we knew details of the $D$-dimensional theory. However, we do not know 
it. Thus we shall determine the constant $A$ phenomenologically. We require 
that the total potential $U_{0}(\sigma)+V_{1loop}(\sigma)$ of the field 
$\sigma$ has an extremum at $\sigma =0$ (or at $b=b_{0}$, where $b_{0}$ 
is the today's value of $b$) and the
extremum (or the $4$-dimensional cosmological constant) is zero. Thus 
the constant $A$ and the renormalized value of the $D$-dimensional 
cosmological constant $\bar{\Lambda}$ can be expressed in terms of 
$b_{0}$, and the total potential of $\sigma$ is 
%============< EQUATION >==============%
%
\begin{equation}
 U_{1}(\sigma) = \alpha\left[
 	\frac{2}{d+2}e^{-2(d+2)\sigma /\sigma_{0}} 
 	+ e^{-d\sigma /\sigma_{0}}
 	- \frac{d+4}{d+2}e^{-(d+2)\sigma /\sigma_{0}}\right],
\end{equation}
%======================================%
where $\alpha$ is a constant defined by
%============< EQUATION >==============%
%
\begin{equation}
 \alpha = \frac{d(d-1)(d+2)}{2(d+4)\kappa^2b_{0}^2}.
\end{equation}
%======================================%
This form of the total potential of the field $\sigma$ is used in many 
literature~\cite{Potential-U1}. 
Its second derivative at $\sigma =0$ is given by
%============< EQUATION >==============%
%
\begin{equation}
 U_{1}''(0) = \frac{2(d-1)}{b_{0}^2},
\end{equation}
%======================================%
and is positive for $d\ge 2$. Thus, for $d\ge 2$, the potential $U_{1}$ 
has a local minimum at $\sigma =0$. Finally the $4$-dimensional action 
for the fields $g_{\mu\nu}$, $\sigma$ and $\phi_{lm}$ in this model 
is 
%============< EQUATION >==============%
%
\begin{eqnarray}
 I & = & \int d^4x\sqrt{-g}\left[\frac{1}{2\kappa^2}R
	-\frac{1}{2}g^{\mu\nu}
	\partial_{\mu}\sigma\partial_{\nu}\sigma 
	- U_{1}(\sigma)\right] \nonumber\\
 & & -\frac{1}{2}\sum_{l,m}\int d^4x\sqrt{-g}\left[
 	g^{\mu\nu}\partial_{\mu}\phi_{lm}\partial_{\nu}\phi_{lm} 
 	+ M_{l}^2(\sigma)\phi_{lm}^2\right].
	\label{eqn:total-action1}
\end{eqnarray}
%======================================%

\subsection{Compactification by $S_{d_{1}}\times S_{d_{2}}$}

Next we consider a compactification by a direct product 
$S_{d_{1}}\times S_{d_{2}}$ of a $d_{1}$-dimensional sphere and a 
$d_{2}$-dimensional sphere $(d_{1}+d_{2}=D-4)$. We assume that 
$d_{1}\ge d_{2}\ge 1$. In this case the metric 
tensor $\bar{g}_{MN}$ can be written as
%============< EQUATION >==============%
%
\begin{equation}
 \bar{g}_{MN}dx^Mdx^N = \hat{g}_{\mu\nu}dx^{\mu}dx^{\nu}
	+ b_{1}^2\Omega^{(d_{1})}_{ij}dx^idx^j
	+ b_{2}^2\Omega^{(d_{2})}_{pq}dx^pdx^q,
\end{equation}
%======================================%
where $\hat{g}_{\mu\nu}$ is a $4$-dimensional metric depending only on 
the $4$-dimensional coordinates $x^{\mu}$ $(\mu=0,1,2,3)$, $b_{1}$ and 
$b_{2}$ are radii of the $d_{1}$-dimensional sphere and the 
$d_{2}$-dimensional sphere depending only on the $4$-dimensional coordinates, 
and $\Omega^{(d_{1})}_{ij}dx^idx^j$ and $\Omega^{(d_{2})}_{pq}dx^pdx^q$ 
are line elements of a unit $d_{1}$-sphere and a unit $d_{2}$-sphere, 
respectively. The scalar field $\bar{\phi}$ is expanded as follows: 
%============< EQUATION >==============%
%
\begin{equation}
 \bar{\phi} = b_{10}^{-d_{1}/2}b_{20}^{-d_{2}/2}
	\sum_{l_{1},l_{2},m_{1},m_{2}}
	\phi_{l_{1}l_{2}m_{1}m_{2}}(x^{\mu})
	Y^{(d_{1})}_{l_{1}m_{2}}(x^i) Y^{(d_{2})}_{l_{2}m_{2}}(x^p),
\end{equation}
%======================================%
where $b_{10}$ and $b_{20}$ are today's values of 
$b_{1}$ and $b_{2}$, $Y^{(d_{1})}_{l_{1}m_{1}}$ and 
$Y^{(d_{2})}_{l_{2}m_{2}}$ are real harmonics on the $d_{1}$-sphere and 
the $d_{2}$-sphere, respectively, and $\phi_{l_{1}l_{2}m_{1}m_{2}}(x^{\mu})$ 
is a real function depending only on the $4$-dimensional coordinates 
$x^{\mu}$. The Einstein-Hilbert action and the Klein-Gordon action in
this case are 
%============< EQUATION >==============%
%
\begin{eqnarray}
 I_{EH} & = & \int d^4x\sqrt{-g}\left[\frac{1}{2\kappa^2}R
	-\frac{1}{2}g^{\mu\nu}\partial_{\mu}\sigma_{+}\partial_{\nu}\sigma_{+}
	-\frac{1}{2}g^{\mu\nu}\partial_{\mu}\sigma_{-}\partial_{\nu}\sigma_{-}
	- U_{0}(\sigma_{1},\sigma_{2})\right] ,\nonumber\\
 I_{KG} & = & -\frac{1}{2}\sum_{l_{1},l_{2},m_{1},m_{2}}
 	\int d^4x\sqrt{-g}\left[
 	g^{\mu\nu}\partial_{\mu}\phi_{l_{1}l_{2}m_{1}m_{2}}
 	\partial_{\nu}\phi_{l_{1}l_{2}m_{1}m_{2}}
 	+ M_{l_{1}l_{2}}^2(\sigma_{1},\sigma_{2})
 	\phi_{l_{1}l_{2}m_{1}m_{2}}^2\right],
\end{eqnarray}
%======================================%
where we have performed the conformal transformation~\footnote{
See the comments before (\ref{eqn:conformal-tr}).
}
%============< EQUATION >==============%
%
\begin{equation}
 \hat{g}_{\mu\nu} = 
 	\left(\frac{b_{1}}{b_{10}}\right)^{-d_{1}}
	\left(\frac{b_{2}}{b_{20}}\right)^{-d_{2}}g_{\mu\nu}.
\end{equation}
%======================================%
In the expressions, $\kappa$ is a positive constant defined by 
$\kappa^2=
\bar{\kappa}^2/(2^{d_{1}+d_{2}}b_{10}^{d_{1}}b_{20}^{d_{2}}\pi^2)$, 
$R$ is a Ricci scalar of the $4$-dimensional metric tensor 
$g_{\mu\nu}$. The $4$-dimensional fields $\sigma_{1,2}$ are defined by
%============< EQUATION >==============%
%
\begin{eqnarray}
 \sigma_{1,2} & = & 
 	\sigma_{10,20}\ln\left(\frac{b_{1,2}}{b_{10,20}}\right)
 	,\nonumber\\
 \sigma_{10,20} & = & \sqrt{\frac{d_{1,2}(d_{1,2}+2)}{2\kappa^2}},
\end{eqnarray}
%======================================%
and $\sigma_{\pm}$ are linear combinations of $\sigma_{1,2}$:
%============< EQUATION >==============%
%
\begin{equation}
 \sigma_{\pm} = \sqrt{\frac{1\pm a}{2}}
 	\left(\sigma_{1}\pm\sigma_{2}\right),
\end{equation}
%======================================%
where $a=\sqrt{\frac{d_{1}d_{2}}{(d_{1}+2)(d_{2}+2)}}$. The mass 
$M_{l_{1}l_{2}}(\sigma_{1},\sigma_{2})$ of the $4$-dimensional scalar field 
$\phi_{l_{1}l_{2}m_{1}m_{2}}$ is given by
%============< EQUATION >==============%
%
\begin{equation}
 M_{l_{1}l_{2}}^2(\sigma_{1},\sigma_{2}) = 
 	e^{-d_{1}\sigma_{1}/\sigma_{10}-d_{2}\sigma_{2}/\sigma_{20}}
	\left[
	\frac{l_{1}(l_{1}+d_{1}-1)}{b_{10}^2}
	e^{-2\sigma_{1}/\sigma_{10}}
 	+ \frac{l_{2}(l_{2}+d_{2}-1)}{b_{20}^2}
 	e^{-2\sigma_{2}/\sigma_{20}}\right]
 	,\label{eqn:mass2}
\end{equation}
%======================================%
and the potential $U_{0}(\sigma_{1},\sigma_{2})$ is
%============< EQUATION >==============%
%
\begin{equation}
 U_{0}(\sigma_{1},\sigma_{2}) = 
 	e^{-d_{1}\sigma_{1}/\sigma_{10}-d_{2}\sigma_{2}/\sigma_{20}}
 	\left[\frac{\bar{\Lambda}}{\kappa^2}
 	-\frac{d_{1}(d_{1}-1)}{2\kappa^2b_{10}^2}
 	e^{-2\sigma_{1}/\sigma_{10}}
 	-\frac{d_{2}(d_{2}-1)}{2\kappa^2b_{20}^2}
 	e^{-2\sigma_{2}/\sigma_{20}}\right].
\end{equation}
%======================================%
Here we mention that a mode with non-zero $l_1$ or non-zero $l_2$ has
a mass of order $1/b_{10}$ or $1/b_{20}$ and is called a Kaluza-Klein
mode.

As in the case of the compactification by $S_{d}$, it can be easily 
confirmed that $U_{0}(\sigma_{1},\sigma_{2})$ has no local minimum, 
and that there is no stable compactification by 
$S_{d_{1}}\times S_{d_{2}}$ if the model is not modified any more. 
In order to stabilize the potential we consider a $1$-loop effective 
action (an action for so-called Casimir effects) 
contributed by all $4$-dimensional matter fields. The $1$-loop effective 
action is of the following form \cite{Candelas&Weinberg}: 
%============< EQUATION >==============%
%
\begin{equation}
 I_{1loop} = -\int d^4x\sqrt{-g}V_{1loop}(\sigma_{1},\sigma_{2}),
\end{equation}
%======================================%
where $V_{1loop}$ is an effective potential, which in turn is a sum over 
all contributing fields~\footnote{
See the footnote soon after (\ref{eqn:1loop-I-V}).
}.
%============< EQUATION >==============%
%
\begin{equation}
 V_{1loop}(\sigma_{1},\sigma_{2}) = \sum_{i}c_{i}V(M_{i}),
\end{equation}
%======================================%
where $i$ distinguishes $4$-dimensional fields contributing to the 
$1$-loop effective action, $M_{i}$ is a mass of the field specified by 
$i$, and $c_{i}$ is a numerical factor depending on what spin the field 
$i$ has~\footnote{
See the footnote soon after (\ref{eqn:1loop-V}).
}.
In the expression the function $V(M)$ is given by (\ref{eqn:VM}). 
Since, in the present model, squared masses of $4$-dimensional fields 
deduced from the $D$-dimensional massless theory is a sum of a term 
proportional to 
$e^{-(d_{1}+2)\sigma_{1}/\sigma_{10}-d_{2}\sigma_{2}/\sigma_{20}}$ 
and a term proportional to 
$e^{-d_{1}\sigma_{1}/\sigma_{10}-(d_{2}+2)\sigma_{2}/\sigma_{20}}$ 
as the squared mass (\ref{eqn:mass2}) of the 
$\phi_{lm}$ fields, the $1$-loop effective action is of the following 
form in general. 
%============< EQUATION >==============%
%
\begin{equation}
 V_{1loop}(\sigma_{1},\sigma_{2}) = 
 	e^{-2d_{1}\sigma_{1}/\sigma_{10}-2d_{2}\sigma_{2}/\sigma_{20}}
 	\left(A_{11}e^{-4\sigma_{1}/\sigma_{10}}
 	+2A_{12}e^{-2\sigma_{1}/\sigma_{10}-2\sigma_{2}/\sigma_{20}}
 	+A_{22}e^{-4\sigma_{2}/\sigma_{20}}\right),
\end{equation}
%======================================%
where $A_{11}$, $A_{12}$ and $A_{22}$ are constants. We require that the 
total potential 
$U_{0}(\sigma_{1},\sigma_{2})+V_{1loop}(\sigma_{1},\sigma_{2})$ of the 
fields $\sigma_{1}$ and $\sigma_{2}$ has an extremum at 
$\sigma_{1}=\sigma_{2}=0$ (or at $b_{1}=b_{10}$, $b_{2}=b_{20}$, 
where $b_{10}$ and $b_{20}$ are the today's values of $b_{1}$ and 
$b_{2}$, respectively) and the 
extremum (or the $4$-dimensional cosmological constant) is 
zero. Thus $A_{11}$, $A_{22}$ and $\bar{\Lambda}$ can be expressed in 
terms of $b_{10}$, $b_{20}$ and the unknown constant $A_{12}$, and the 
total potential of $\sigma_{1}$ and $\sigma_{2}$ is 
%============< EQUATION >==============%
%
\begin{eqnarray}
 U_{1}(\sigma_{1},\sigma_{2}) & = &
	\alpha_1U_1^{(1)}(\sigma_{1},\sigma_{2}) +
	\alpha_2U_1^{(2)}(\sigma_{1},\sigma_{2})	\nonumber\\
 & & 
	-A_{12}e^{-2d_{1}\sigma_{1}/\sigma_{10}-2d_{2}\sigma_{2}/\sigma_{20}}
	\left( e^{-2\sigma_{1}/\sigma_{10}}
		-e^{-2\sigma_{2}/\sigma_{20}}\right)^2,
\end{eqnarray}
%======================================%
where the constants $\alpha_{1,2}$ and the functions
$U_1^{(1,2)}(\sigma_{1},\sigma_{2})$ are defined by
%============< EQUATION >==============%
%
\begin{eqnarray}
 \alpha_{1} & = & \frac{d_1(d_1-1)(d_1+d_2+2)}
	{2(d_1+d_2+4)\kappa^2b_{10}^2},\nonumber\\
 \alpha_{2} & = & \frac{d_2(d_2-1)(d_1+d_2+2)}
	{2(d_1+d_2+4)\kappa^2b_{20}^2},\nonumber\\
 U_1^{(1)}(\sigma_{1},\sigma_{2}) & = &
	\frac{d_2+4}{2(d_1+d_2+2)}
	e^{-2(d_1+2)\sigma_1/\sigma_{10}-2d_2\sigma_2/\sigma_{20}}
	+e^{-d_1\sigma_1/\sigma_{10}-d_2\sigma_2/\sigma_{20}}
	\nonumber\\
 & & -\frac{d_1+d_2+4}{d_1+d_2+2}
	e^{-(d_1+2)\sigma_1/\sigma_{10}-d_2\sigma_2/\sigma_{20}}
	\nonumber\\
 & & -\frac{d_2}{2(d_1+d_2+2)}
	e^{-2d_1\sigma_1/\sigma_{10}-2(d_2+2)\sigma_2/\sigma_{20}}
	,\nonumber\\
 U_1^{(2)}(\sigma_{1},\sigma_{2}) & = &
	\frac{d_1+4}{2(d_1+d_2+2)}
	e^{-2d_1\sigma_1/\sigma_{10}-2(d_2+2)\sigma_2/\sigma_{20}}
	+e^{-d_1\sigma_1/\sigma_{10}-d_2\sigma_2/\sigma_{20}}
	\nonumber\\
 & & -\frac{d_1+d_2+4}{d_1+d_2+2}
	e^{-d_1\sigma_1/\sigma_{10}-(d_2+2)\sigma_2/\sigma_{20}}
	\nonumber\\
 & & -\frac{d_1}{2(d_1+d_2+2)}
	e^{-2(d_1+2)\sigma_1/\sigma_{10}-2d_2\sigma_2/\sigma_{20}}.
\end{eqnarray}
%======================================%
Finally the $4$-dimensional action for the fields $g_{\mu\nu}$, 
$\sigma_{1}$, $\sigma_{2}$ and $\phi_{l_{1}l_{2}m_{1}m_{2}}$ in this model 
is 
%============< EQUATION >==============%
%
\begin{eqnarray}
 I & = & \int d^4x\sqrt{-g}\left[\frac{1}{2\kappa^2}R
	-\frac{1}{2}g^{\mu\nu}
	\partial_{\mu}\sigma_{+}\partial_{\nu}\sigma_{+} 
	-\frac{1}{2}g^{\mu\nu}
	\partial_{\mu}\sigma_{-}\partial_{\nu}\sigma_{-} 
	- U_{1}(\sigma_{1},\sigma_{2})\right] \nonumber\\
 & & -\frac{1}{2}\sum_{l_{1},l_{2},m_{1},m_{2}}\int d^4x\sqrt{-g}
 	\left[g^{\mu\nu}\partial_{\mu}\phi_{l_{1}l_{2}m_{1}m_{2}}
 	\partial_{\nu}\phi_{l_{1}l_{2}m_{1}m_{2}}
 	+ M_{l_{1}l_{2}}^2(\sigma)
 	\phi_{l_{1}l_{2}m_{1}m_{2}}^2\right].\label{eqn:action2}
\end{eqnarray}
%======================================%

Unfortunately, the potential $U_{1}(\sigma_{1},\sigma_{2})$ depends on 
the unknown constant $A_{12}$. In order to make our arguments 
independent of the value of $A_{12}$ we take the following ansatz:
%============< EQUATION >==============%
%
\begin{equation}
 \frac{b_{1}}{b_{10}} = \frac{b_{2}}{b_{20}}
 	.\label{eqn:b1-b2-ansatz}
\end{equation}
%======================================%
Physically, this condition means that the shape of the extra 
$(D-4)$-dimensional space $S_{d_{1}}\times S_{d_{2}}$ is fixed and only 
its volume changes. In this case let us introduce a field $\sigma$ by
%============< EQUATION >==============%
%
\begin{eqnarray}
 \sigma /\sigma_{0} & = & \sigma_{1}/\sigma_{10}
	= \sigma_{2}/\sigma_{20},\nonumber\\
 \sigma_{0} & = & 
	\sqrt{\frac{(d_{1}+d_{2})(d_{1}+d_{2}+2)}{2\kappa^2}}.
\end{eqnarray}
%======================================%
The action (\ref{eqn:action2}) is simplified as
%============< EQUATION >==============%
%
\begin{eqnarray}
 I & = & \int d^4x\sqrt{-g}\left[\frac{1}{2\kappa^2}R
	-\frac{1}{2}g^{\mu\nu}\partial_{\mu}\sigma\partial_{\nu}\sigma
	- U_{1}(\sigma)\right] \nonumber\\
 & & -\frac{1}{2}\sum_{l_{1},l_{2},m_{1},m_{2}}
 	\int d^4x\sqrt{-g}\left[
 	g^{\mu\nu}\partial_{\mu}\phi_{l_{1}l_{2}m_{1}m_{2}}
 	\partial_{\nu}\phi_{l_{1}l_{2}m_{1}m_{2}}
 	+ M_{l_{1}l_{2}}^2(\sigma)\phi_{l_{1}l_{2}m_{1}m_{2}}^2\right]
	,\label{eqn:total-action2}
\end{eqnarray}
%======================================%
where the mass $M_{l_{1}l_{2}}(\sigma)$ of the Kaluza-Klein mode
$\phi_{l_{1}l_{2}m_{1}m_{2}}$ is given by
%============< EQUATION >==============%
%
\begin{equation}
 M_{l_{1}l_{2}}^2(\sigma) = \left[
 	\frac{l_{1}(l_{1}+d_{1}-1)}{b_{10}^2} + 
 	\frac{l_{2}(l_{2}+d_{2}-1)}{b_{20}^2}\right]
 	e^{-(d_{1}+d_{2}+2)\sigma /\sigma_{0}},
\end{equation}
%======================================%
and the potential $U_{1}(\sigma)$ of the $\sigma$ field is
%============< EQUATION >==============%
%
\begin{eqnarray}
 U_{1}(\sigma) & = & (\alpha_1+\alpha_2)\left[
	\frac{2}{d_1+d_2+2}e^{-2(d_1+d_2+2)\sigma/\sigma_0}
	+e^{-(d_1+d_2)\sigma/\sigma_0}\right.
	\nonumber\\
 & & \left.-\frac{d_1+d_2+4}{d_1+d_2+2}
	e^{-(d_1+d_2+2)\sigma/\sigma_0}\right].
\end{eqnarray}
%======================================%
Note that the potential $U_{1}(\sigma)$ is independent of the unknown 
constant $A_{12}$.The second derivative of $U_{1}(\sigma)$ at $\sigma
=0$ is 
%============< EQUATION >==============%
%
\begin{equation}
 U_{1}''(0) = \frac{2}{d_1+d_2}\left[
	\frac{d_1(d_1-1)}{d_{10}^2}+\frac{d_2(d_2-1)}{d_{20}^2}
	\right].
\end{equation}
%======================================%
Hence, if $d_{1}\ge 2$, the extremum of
$U_{1}(\sigma)$ at $\sigma =0$ is a local minimum and the
compactification is stable at least locally.

%======================================%
%<<<<<<<<<<<< SECTION III >>>>>>>>>>>>>%p
%======================================%

\section{Parametric resonance}\label{sec:resonance}

In the previous section we have shown that the compactification by
$S_d$ $(d\ge 2)$ is stable, and the compactification by
$S_{d_1}\times S_{d_{2}}$ $(d_1\ge 2)$ is also stable at least in the 
sub-configuration-space defined by (\ref{eqn:b1-b2-ansatz}). Thus, 
we consider the $S_{d}$ 
compactification with $d\ge 2$ and the $S_{d_1}\times S_{d_2}$ 
compactification with $d_{1}\ge 2$.  In this section we judge
whether a catastrophic creation of quanta of the Kaluza-Klein mode 
$\phi_{lm}$ (or $\phi_{l_1l_2m_1m_2}$) can occur due to an 
oscillation of the $\sigma$ field around the local minimum $\sigma
=0$. We treat the metric tensor $g_{\mu\nu}$ and the field $\sigma$
classically, and the fields $\phi_{lm}$ (or $\phi_{l_1l_2m_1m_2}$)
quantum mechanically.

\subsection{The hamiltonians for $\phi_{lm}$ and $\phi_{l_1l_2m_1m_2}$}

We assume that the $4$-dimensional metric $g_{\mu\nu}$ is of the
Friedmann universe and that the field $\sigma$ is homogeneous on
this background. The following equation of motion for the 
field $\sigma$ is derived from the $4$-dimensional action
(\ref{eqn:total-action1}) or (\ref{eqn:total-action2}):
%============< EQUATION >==============%
%
\begin{equation}
 \ddot{\sigma}+3H\dot{\sigma}+U_{1}'(\sigma)=0
	,\label{eqn:eq-of-sigma}
\end{equation}
%======================================%
where $H=\dot{a}/a$ ($a$ is a scale factor of the universe) and the 
overdot denotes derivation with respect to the cosmological time $t$. 
Hereafter we assume that 
%============< EQUATION >==============%
%
\begin{equation}
 H \ll \omega,\label{eqn:H<<omega}
\end{equation}
%======================================%
where $\omega$ is defined by
%============< EQUATION >==============%
%
\begin{equation}
 \omega = \frac{1}{2}\sqrt{U_{1}''(0)}.
\end{equation}
%======================================%
Hence for a small deviation from the local minimum $\sigma =0$ 
($\left|\sigma/\sigma_0\right|\ll 1/D$), a solution $\sigma(t)$ 
of the equation (\ref{eqn:eq-of-sigma}) can be written as
%============< EQUATION >==============%
%
\begin{equation}
 \frac{\sigma(t)}{\sigma_0} = 
	\tilde{\sigma}(t)\cos [ 2\omega(t-t_0)]
	,\label{eqn:sigma-cos}
\end{equation}
%======================================%
where the coefficient $\tilde{\sigma}(t)$
($\left|\tilde{\sigma}(t)\right|\ll 1/D$) is a slowly varying 
function of $t$:  
%============< EQUATION >==============%
%
\begin{equation}
 \frac{\dot{\tilde{\sigma}}}{\tilde{\sigma}} 
	\ll \omega, 
\end{equation}
%======================================%
and $t_0$ is an arbitrary constant. 
In order to determine a time dependence of the function
$\tilde{\sigma}(t)$, we introduce energy density $\rho_{\sigma}$ of
the $\sigma$ field by 
%============< EQUATION >==============%
%
\begin{equation}
 \rho_{\sigma} \equiv 
	\frac{1}{2}\dot{\sigma}^2 
	+ \frac{1}{2}(2\omega)^2\sigma^2.
\end{equation}
%======================================%
From (\ref{eqn:eq-of-sigma}) the following equation is easily derived:
%============< EQUATION >==============%
%
\begin{equation}
 \dot{\rho_{\sigma}} = -3H\dot{\sigma}^2. 
\end{equation}
%======================================%
By averaging this equation from $t-\pi /(2\omega)$ to 
$t+\pi /(2\omega)$, we obtain the following approximate
equation. 
%============< EQUATION >==============%
%
\begin{equation}
 \dot{\rho_{\sigma}} \simeq 
	-\frac{3}{2}H (2\omega)^2(\sigma_0\tilde{\sigma})^2.
	\label{eqn:averaged-rho-eq}
\end{equation}
%======================================%
Since in our situation the value of $\rho_{\sigma}$ is approximately
given by 
%============< EQUATION >==============%
%
\begin{equation}
 \rho_{\sigma} \simeq 
	\frac{1}{2}(2\omega)^2(\sigma_0\tilde{\sigma})^2, 
	\label{eqn:rho-sigma}
\end{equation}
%======================================%
the time dependence of $\tilde{\sigma}$ is derived from the averaged
equation (\ref{eqn:averaged-rho-eq}) as
%============< EQUATION >==============%
%
\begin{equation}
 \tilde{\sigma}(t) \propto a^{-3/2}.\label{eqn:sigma-a}
\end{equation}
%======================================%

In the remaining of this section we investigate a quantum dynamics of
the Kaluza-Klein mode $\phi_{lm}$ (for the $S_d$ case) and
$\phi_{l_1l_2m_1m_2}$ (for the $S_{d_1}\times S_{d_2}$ case) on the
classical background of the Friedmann universe and the field $\sigma$
given by (\ref{eqn:sigma-cos}) and (\ref{eqn:sigma-a}). For this
purpose we first go to the momentum representation by 
%============< EQUATION >==============%
%
\begin{eqnarray}
 \phi_{lm}({\bf x},t) & = & a^{-3/2}
 	\int\frac{\sqrt{2}d^3{\bf k}}{(2\pi)^3}\left[
 	\phi^{(1){\bf k}}_{lm}(t)\cos ({\bf k}\cdot{\bf x})
 	+ \phi^{(2){\bf k}}_{lm}(t)\sin ({\bf k}\cdot{\bf x})
 	\right],\\
 \phi_{l_1l_2m_1m_2}({\bf x},t) & = & a^{-3/2}
 	\int\frac{\sqrt{2}d^3{\bf k}}{(2\pi)^3}\left[
 	\phi^{(1){\bf k}}_{l_1l_2m_1m_2}(t)\cos({\bf k}\cdot{\bf x})
 	+ \phi^{(2){\bf k}}_{l_1l_2m_1m_2}(t)\sin({\bf k}\cdot{\bf x})
 	\right],
\end{eqnarray}
%======================================%
where $\phi^{(1,2){\bf k}}_{lm}(t)$ and 
$\phi^{(1,2){\bf k}}_{l_1l_2m_1m_2}(t)$ are real functions of the 
cosmological time $t$, the vector ${\bf x}$ denotes the comoving
coordinates~\footnote{
To obtain the representation, for concreteness, we assume that the
Friedmann universe is flat one. If we assume the open or closed
Friedmann universe then the corresponding representation is
changed. However, the resulting partial hamiltonians are of the same
form as (\ref{eqn:hamiltonian-phi}), provided that ${\bf k}^2$ is
replaced by the corresponding eigen value. 
}. Next we obtain the Hamiltonian
corresponding to the time $t$ from the action
(\ref{eqn:total-action1}) and (\ref{eqn:total-action2}). The result is
summarized as follows: 
%============< EQUATION >==============%
%
\begin{equation}
 H_{S_d} = 
 	\sum_{l,m}\int d{\bf k}\left( 
 	H^{(1){\bf k}}_{lm} + H^{(2){\bf k}}_{lm}\right)
\end{equation}
%======================================%
for the $S_d$ case, and 
%============< EQUATION >==============%
%
\begin{equation}
 H_{S_{d_1}\times S_{d_2}} = 
 	\sum_{l_{1},l_{2},m_{1},m_{2}}\int d{\bf k}\left(
 	H^{(1){\bf k}}_{l_{1}l_{2}m_{1}m_{2}} + 
 	H^{(2){\bf k}}_{l_{1}l_{2}m_{1}m_{2}}\right)
\end{equation}
%======================================%
for the $S_{d_1}\times S_{d_2}$ case. In the expressions the partial 
hamiltonians $H^{(1,2){\bf k}}_{lm}$ and 
$H^{(1,2){\bf k}}_{l_{1}l_{2}m_{1}m_{2}}$ are of the following form. 
%============< EQUATION >==============%
%
\begin{equation}
 H = \frac{1}{2}\left[P^2 + \Omega^2(t)Q^2\right]
 	,\label{eqn:hamiltonian-phi}
\end{equation}
%======================================%
where $Q$ denotes $\phi^{(1,2){\bf k}}_{lm}$ for 
$H^{(1,2){\bf k}}_{lm}$ and $\phi^{(1,2){\bf k}}_{l_1l_2m_1m_2}$ 
for $H^{(1,2){\bf k}}_{l_{1}l_{2}m_{1}m_{2}}$, and $P$ denotes its
conjugate momentum. The function $\Omega(t)$ is a positive function of 
$t$ given by~\footnote{
We can set $t_0=0$ without loss of generality.
}
%============< EQUATION >==============%
%
\begin{equation}
 \Omega^2(t) = \omega^2\left[A+Be^{-\epsilon\cos{2\omega t}}\right]
 	,\label{eqn:Omega-phi}
\end{equation}
%======================================%
where the parameters $A$, $B$, $\epsilon$ and $\omega$ are given by
%============< EQUATION >==============%
%
\begin{eqnarray}
 A & = & \frac{2b_0^2\left({\bf k}/a\right)^2}{d-1} + \Delta A
 	,\nonumber\\
 B & = & \frac{2}{d-1}l(l+d-1),\nonumber\\
 \epsilon & = & (d+2)\tilde{\sigma},\nonumber\\
 \omega  & = & \frac{1}{b_{0}}\sqrt{\frac{d-1}{2}}\ \ 
	\label{eqn:pareters-Sd}
\end{eqnarray}
%======================================%
for the $S_d$ case~\footnote{
If $l=0$, then $B=0$ and the parametric resonance does not arise. 
Thus, hereafter we consider a mode with $l\ge 1$. 
}, 
and
%============< EQUATION >==============%
%
\begin{eqnarray}
 A & = & \frac{2(d_1+d_2)b_{10}^2({\bf k}/a)^2}
	{d_1(d_1-1)+d_2(d_2-1)(b_{10}/b_{20})^2}
	+\Delta A ,\nonumber\\
 B & = & \frac{2(d_1+d_2)
	\left[ l_1(l_1+d_1-1)+l_2(l_2+d_2-1)(b_{10}/b_{20})^2\right]}
	{d_1(d_1-1)+d_2(d_2-1)(b_{10}/b_{20})^2},\nonumber\\
 \epsilon & = & (d_1+d_2+2)\tilde{\sigma},\nonumber\\
 \omega  & = & \sqrt{\frac{1}{2(d_1+d_2)}\left[
	\frac{d_1(d_1-1)}{b_{10}^2}+\frac{d_2(d_2-1)}{b_{20}^2}
	\right]}
	\label{eqn:pareters-Sd1*Sd2}
\end{eqnarray}
%======================================%
for the $S_{d_1}\times S_{d_2}$ case~\footnote{
If $l_{1}=l_{2}=0$, then $B=0$ and the parametric resonance does not 
arise. Thus, hereafter we consider a mode with $l_{1}\ge 1$ or 
$l_{2}\ge 1$. 
}, 
respectively. In both expressions of $A$, the term $\Delta A$ 
denotes 
%============< EQUATION >==============%
%
\begin{equation}
 \Delta A =	-\omega^{-2}
 	\left(\frac{9}{4}H^2+\frac{3}{2}\dot{H}\right).
\end{equation}
%======================================%

In the remaining of this paper we consider the case when the condition 
(\ref{eqn:H<<omega}) and the following condition hold. 
%============< EQUATION >==============%
%
\begin{equation}
 \dot{H} \ll \omega^2.
\end{equation}
%======================================%
Hence we can neglect the term $\Delta A$ in $A$. In general the
parameters $A$ and $\epsilon$ appearing in the function $\Omega$ have
time dependence even when we neglect the term $\Delta A$. However, 
since $A\propto a^{-2}$ and $\epsilon\propto a^{-3/2}$, it is shown from  
(\ref{eqn:H<<omega}) that their change is extremely slow:
%============< EQUATION >==============%
%
\begin{eqnarray}
 \left|\frac{\dot{A}}{A}\right| & = & 2H\ll \omega
 	,\nonumber\\ 
 \left|\frac{\dot{\epsilon}}{\epsilon}\right| & = & \frac{3}{2}H
 	\ll \omega.
\end{eqnarray}
%======================================%

\subsection{The resonance band and the number of created quanta}

A system with a hamiltonian like (\ref{eqn:hamiltonian-phi}), when it is 
canonically quantized, has the property that a catastrophic creation of 
quanta (parametric resonance) can occur if the parameters $A$, $B$ and 
$\epsilon$ is in a band in their parameter space. Our task now is to
derive the band and a number of created quanta of the Kaluza-Klein
mode $\phi_{lm}$ or $\phi_{l_{1}l_{2}m_{1}m_{2}}$. First we expand  
(\ref{eqn:Omega-phi}) both in power of $\epsilon$ and in Fourier 
series. For this purpose we use the identity 
%============< EQUATION >==============%
%
\begin{equation}
 e^{-\epsilon\cos{2z}} = I_{0}(\epsilon) +
 	2\sum_{n=1}^{\infty}(-1)^nI_{n}(\epsilon)\cos{2nz},
\end{equation}
%======================================%
where $I_{n}(x)$ is the modified Bessel function and is expanded as
%============< EQUATION >==============%
%
\begin{equation}
 I_{n}(x) = \sum_{s=0}^{\infty}\frac{(x/2)^{n+2s}}{s!(n+s)!}. 
\end{equation}
%======================================%
The obtained expression is 
%============< EQUATION >==============%
%
\begin{equation}
 \Omega^2(t) = \omega^2\left[h-
   \sum_{s=1}^{\infty}\sum_{n=-\infty}^{\infty}
   \epsilon^s g^{(s)}_{n}e^{inz}\right]
\end{equation}
%======================================%
where $z=\omega t$. In the expression the constants $h$ and $g^{(s)}_{n}$ 
are given by 
%============< EQUATION >==============%
%
\begin{eqnarray}
 h & = & A + B I_{0}(\epsilon),\nonumber\\
 g_{\pm 2n}^{(n+2s)} & = & \frac{(-1)^{n+1}B}{2^{n+2s}s!(n+s)!},
\end{eqnarray}
%======================================%
where $n=1,2,\cdots$; $s=0,1,\cdots$, and other $g_{n}^{(s)}$ are all 
zero. For example,
%============< EQUATION >==============%
%
\begin{eqnarray}
 g_{\pm 2}^{(1)} & = & \frac{B}{2},\ \ 
 g_{n}^{(1)} = 0\ \ (n\ne\pm 2),\nonumber\\
 g_{\pm 4}^{(2)} & = & -\frac{B}{8},\ \ 
 g_{n}^{(2)} = 0\ \ (n\ne\pm 4).
\end{eqnarray}
%======================================%

Next let us introduce a parameter $\Delta$ by
%============< EQUATION >==============%
%
\begin{equation}
 h = \left( \frac{p}{q}\right)^2 + \epsilon\Delta,
\end{equation}
%======================================%
where $p$ and $q$ are mutually prime positive integers. 
In appendix \ref{app:Floquet} and \ref{app:number-quanta} it is shown 
that the parametric resonance does occur if the parameter $\Delta$ is 
in the following band: 
%============< EQUATION >==============%
%
\begin{equation}
 \Delta_{0}-|\delta| < \Delta < \Delta_{0}+|\delta|	,
\end{equation}
%======================================%
where $\Delta_{0}$ is a combination of the coefficients $g_{n}^{(s)}$ 
being of order $O(\epsilon)$, and $\delta$ is defined by 
%============< EQUATION >==============%
%
\begin{equation}
 \delta  = g^{(1)}_{2p/q} +\epsilon\left[\sum_{n\ne 0,2p/q}
 	\frac{g^{(1)}_{n} g^{(1)}_{2p/q-n}}{n(2p/q-n)}
 	+g^{(2)}_{2p/q}\right]. 
\end{equation}
%======================================%
Thus the existence of non-trivial interval of the resonance band 
requires that $\delta\ne 0$, which in turn requires that $p/q$ is $1$ 
or $2$. Physically the first resonance corresponds to a decay of a 
quantum of the field $\sigma$ with energy $2\omega$ into two quanta 
of the Kaluza-Klein mode with each energy $\omega$, and the second 
resonance corresponds to a decay of a quantum of $\sigma$ into one 
quantum of the Kaluza-Klein mode with energy $2\omega$. The parameters 
$\Delta_{0}$ and $\delta$ in our case are given as follows:  
%============< EQUATION >==============%
%
\begin{eqnarray}
 \Delta_{0} & = & -\frac{B^2}{32}\epsilon,\nonumber\\
 \delta & = & \frac{B}{2}	\label{eqn:delta-Sd}
\end{eqnarray}
%======================================%
for the $p/q=1$ resonance; and 
%============< EQUATION >==============%
%
\begin{eqnarray}
 \Delta_{0} & = & \frac{B^2}{24}\epsilon,\nonumber\\
 \delta & = & \frac{B(B-2)}{16}\epsilon
 \label{eqn:delta-Sd1*S1}
\end{eqnarray}
%======================================%
for the $p/q=2$ resonance. Next the Floquet index $\mu_+$, which 
represents a `strength' of the resonance, is given by the formula 
(\ref{eqn:mu}) and can be expressed as 
%============< EQUATION >==============%
%
\begin{equation}
 \mu_+ = \frac{|\epsilon\delta |}{2p/q}
 	\sqrt{1-x^2}\left( 1+O(\epsilon)\right),
	\label{eqn:mu+}
\end{equation}
%======================================%
where $x$ is defined by $ x=(\Delta-\Delta_{0})/\delta$.

As shown in appendix \ref{app:number-quanta}, by using the form of $\mu_{+}$, 
a number of excited quanta is given by the following formula:
%============< EQUATION >==============%
%
\begin{equation}
 N(t) \simeq \sinh^2\left(\int\mu_+dz\right).
\end{equation}
%======================================%
We now estimate the integral in this formula. If $B<(p/q)^2$ then 
%============< EQUATION >==============%
%
\begin{equation}
 \lim_{a\to +0}x=\pm\infty,\ \ \lim_{a\to\infty}x=\mp\infty,
\end{equation}
%======================================%
and thus we can use the following approximation: 
%============< EQUATION >==============%
%
\begin{equation}
 \int\mu_+dz \simeq  
	\frac{|\epsilon\delta |}{2p/q}
 	\left|\frac{dz}{dx}\right|_{\Delta =\Delta_{0}}
 	\int^{1}_{-1}\sqrt{1-x^{2}}dx
 = \frac{|\epsilon\delta |\pi}{4p/q}
 	\left|\frac{dz}{dx}\right|_{\Delta =\Delta_{0}}.
	\label{eqn:mu-dz}
\end{equation}
%======================================%
The quantity $|dz/dx|_{\Delta =\Delta_{0}}$ can be calculated from
(\ref{eqn:delta-Sd}) or (\ref{eqn:delta-Sd1*S1}) 
as follows~\footnote{ 
To obtain the expressions we have used the fact that $A\propto a^{-2}$
and that $\epsilon\propto a^{-3/2}$. 
}: 
%============< EQUATION >==============%
%
\begin{equation}
 \left|\frac{dz}{dx}\right|_{\Delta =\Delta_{0}} = 
 	\left|\frac{B}{4(B-1)}\right|\cdot
 	\frac{|\epsilon |}{H\omega^{-1}}
 	\left( 1+O(\epsilon^2)\right)
\end{equation}
%======================================%
for the $p/q=1$ resonance with $B<1$; and
%============< EQUATION >==============%
%
\begin{equation}
 \left|\frac{dz}{dx}\right|_{\Delta =\Delta_{0}} = 
 	\left|\frac{B(B-2)}{32(B-4)}\right|
 	\cdot\frac{\epsilon^2}{H\omega^{-1}}
 	\left( 1+O(\epsilon^2)\right)
\end{equation}
%======================================%
for the $p/q=2$ resonance with $B<4$.
If $B=1$ then $x$ for the $p/q=1$ resonance satisfies 
%============< EQUATION >==============%
%
\begin{equation}
 \lim_{a\to\infty}x=0,
\end{equation}
%======================================%
and thus the integral is estimated as 
%============< EQUATION >==============%
%
\begin{equation}
 \int\mu_+dz \simeq 
	\frac{\omega}{4}\int_{t_*}^{\infty}\epsilon dt,
\end{equation}
%======================================%
where $t_*$ is the time when the mode enters the $p/q=1$ resonance
band. If $B=4$ then $x$ for the $p/q=2$ resonance satisfies
%============< EQUATION >==============%
%
\begin{equation}
 \lim_{a\to\infty}x=\infty,
\end{equation}
%======================================%
and thus the integral $\int\mu_+dz$ is given by (\ref{eqn:mu-dz}) at
an order estimate, where $|dz/dx|_{\Delta =\Delta_{0}}$ is given by 
%============< EQUATION >==============%
%
\begin{equation}
 \left|\frac{dz}{dx}\right|_{\Delta =\Delta_{0}} = 
 	\frac{3}{2}\cdot\frac{1}{H\omega^{-1}}
 	\left( 1+O(\epsilon^2)\right).
\end{equation}
%======================================%
Therefore the number of created quanta by the $p/q=1$ resonance is
given by 
%============< EQUATION >==============%
%
\begin{equation}
 N \simeq \sinh^2\left[\frac{B^2\pi}{2^5(B-1)}\cdot
	\frac{\epsilon^2}{H\omega^{-1}}\right]\label{eqn:N1}
\end{equation}
%======================================%
for $B<1$; and 
%============< EQUATION >==============%
%
\begin{equation}
 N \simeq \sinh^2\left[\frac{\omega}{4}
	\int_{t_*}^{\infty}\epsilon dt\right]\label{eqn:N1'}
\end{equation}
%======================================%
for $B=1$. Note that the integral in the expression (\ref{eqn:N1'})
can be performed when a time dependence of the scale factor $a$ is 
specified. The number of created quanta by the $p/q=2$ resonance is 
%============< EQUATION >==============%
%
\begin{equation}
 N \simeq \sinh^2\left[\frac{B^2(B-2)^2\pi}{2^{12}(B-4)}\cdot
	\frac{\epsilon^4}{H\omega^{-1}}\right]\label{eqn:N2}
\end{equation}
%======================================%
for $B<4$; and 
%============< EQUATION >==============%
%
\begin{equation}
 N \simeq \sinh^2\left[ O(1)\cdot
	\frac{\epsilon^2}{H\omega^{-1}}\right]\label{eqn:N2'}
\end{equation}
%======================================%
for $B=4$.

Finally we mention that in order for the parametric resonance to be
efficient to produce the quanta of the Kaluza-Klein mode the following 
condition is necessary and sufficient: 
%============< EQUATION >==============%
%
\begin{equation}
 N\gg 1,\ \ \Gamma\gtrsim 3H,
\end{equation}
%======================================%
where $\Gamma$ is a creation rate of the quanta at the center of the
resonance band given by
%============< EQUATION >==============%
%
\begin{equation}
 \Gamma = 2\omega\left.\mu_+\right|_{x=0}.
\end{equation}
%======================================%
Using the formula (\ref{eqn:mu+}), $\Gamma$ can be written explicitly
as follows:
%============< EQUATION >==============%
%
\begin{equation}
 \Gamma = \frac{B}{2}\epsilon\omega	\label{eqn:Gamma1}
\end{equation}
%======================================%
for $B\le 1$; and 
%============< EQUATION >==============%
%
\begin{equation}
 \Gamma = \frac{B|B-2|}{32}\epsilon^2\omega	\label{eqn:Gamma2}
\end{equation}
%======================================%
for $B\le 4$.

\subsection{The $S_{d}$ case}

We now apply the results obtained in the previous subsection to each
model of the compactification. First, in this subsection we consider
the $S_{d}$ case with $d\ge 2$. In this case it can be seen from
(\ref{eqn:pareters-Sd}) that $B>1$ for $l\ge 1$~\footnote{
See the footnote soon after (\ref{eqn:pareters-Sd}).
}.
Hence the resonance of $p/q=1$ does not occur. On the other hand, the 
resonance of $p/q=2$ can occur for $l=1$ since $B|_{l=1}=4$ for $d=2$
and $2<B|_{l=1}<4$ for $d\ge 3$. For $l\ge 2$, the resonance of
$p/q=2$ does not arise since $B>4$.

For $d=2$ and $l=1$, the number $N$ of created quanta and the creation 
rate $\Gamma$ are given by (\ref{eqn:N2'}) and (\ref{eqn:Gamma2}) with
$B=4$. Hence the parametric resonance is efficient to produce the
quanta of the corresponding Kaluza-Klein mode $\phi_{1m}$ when
%============< EQUATION >==============%
%
\begin{equation}
 \frac{\rho_{\sigma}}{\rho_0} 
	\gtrsim O(1)\cdot\frac{\omega}{H}. \label{eqn:rho-cond1}
\end{equation}
%======================================%
where $\rho_{\sigma}$ is energy density of the oscillation of
$\sigma$, which is approximately given by (\ref{eqn:rho-sigma}), and
$\rho_0$ is so called critical density of the universe defined by 
%============< EQUATION >==============%
%
\begin{equation}
 \rho_0 = \frac{3H^2}{\kappa^2}.
\end{equation}
%======================================%

For $d\ge 3$ and $l=1$, $N$ and $\Gamma$ are given by (\ref{eqn:N2})
and (\ref{eqn:Gamma2}) with $B=2d/(d-1)$.
Hence the parametric resonance is efficient to produce the quanta of
the corresponding Kaluza-Klein mode $\phi_{1m}$ when
%============< EQUATION >==============%
%
\begin{equation}
 \frac{\rho_{\sigma}}{\rho_0} \gtrsim 
	O(1)\cdot\left(\frac{\omega}{H}\right)^{3/2}.\label{eqn:rho-cond2}
\end{equation}
%======================================%

Note that, because of the assumption (\ref{eqn:H<<omega}), the 
condition (\ref{eqn:rho-cond1}) or (\ref{eqn:rho-cond2}) requires 
%============< EQUATION >==============%
%
\begin{equation}
 \frac{\rho_{\sigma}}{\rho_0} \gg 1.
\end{equation}
%======================================%
Thus in our situation ($H\ll\omega$ and $|\sigma /\sigma_{0}\ll1/D$) 
we can conclude that the parametric resonance do not overproduce the 
quanta of the Kaluza-Klein mode in this model.

\subsection{The $S_{d_{1}}\times S_{d_2}$ case}

Following the $S_d$ case, we consider the $S_{d_{1}}\times S_{d_2}$
case. In this model the parameter $B$ is given by
(\ref{eqn:pareters-Sd1*Sd2}).

For $l_1$ and $l_2$ satisfying $B<1$, the number $N$ of created quanta
of the corresponding mode and the creation rate $\Gamma$ are given by
(\ref{eqn:N1}) and (\ref{eqn:Gamma1}), respectively. Hence, if $B<1$ 
and $|B-1|=O(1)$, the parametric resonance is efficient to produce 
the quanta of the Kaluza-Klein mode $\phi_{l_1l_2m_1m_2}$ only when 
%============< EQUATION >==============%
%
\begin{equation}
 \frac{\rho_{\sigma}}{\rho_0} \gtrsim O(1)\cdot\frac{\omega}{H}.
\end{equation}
%======================================%
If $B<1$ and $|B-1|=O(\epsilon )$ then the condition for efficient
production becomes 
%============< EQUATION >==============%
%
\begin{equation}
 \frac{\rho_{\sigma}}{\rho_0} \gtrsim O(1).
\end{equation}
%======================================%

For $l_1$ and $l_2$ satisfying $B=1$, $N$ and $\Gamma$ are given by
(\ref{eqn:N1'}) and (\ref{eqn:Gamma1}) with $B=1$. 
In order to estimate the integral in the expression of $N$ we have to
specify a time dependence of the scale factor $a$. 
If $a\propto e^{Ht}$ then 
%============< EQUATION >==============%
%
\begin{equation}
 \int_{t_*}^{\infty}\epsilon dt= 
	\frac{2}{3}\cdot\left.\frac{\epsilon}{H}\right|_{t=t_*}.
\end{equation}
%======================================%
If $a\propto t^n$ ($n>2/3$) then 
%============< EQUATION >==============%
%
\begin{equation}
 \int_{t_*}^{\infty}\epsilon dt= \frac{2n}{3n-2}\cdot
	\left.\frac{\epsilon}{H}\right|_{t=t_*}.
\end{equation}
%======================================%
If $a\propto t^n$ ($n\le 2/3$) then 
%============< EQUATION >==============%
%
\begin{equation}
 \int_{t_*}^{\infty}\epsilon dt= \infty.
\end{equation}
%======================================%
Anyway, the parametric resonance is efficient to produce the quanta of
the corresponding Kaluza-Klein mode $\phi_{l_1l_2m_1m_2}$ when 
%============< EQUATION >==============%
%
\begin{equation}
 \frac{\rho_{\sigma}}{\rho_0} \gtrsim O(1).
\end{equation}
%======================================%

For $l_1$ and $l_2$ satisfying $1<B<2$ or $2<B<4$, $N$ and $\Gamma$
are given by (\ref{eqn:N2}) and (\ref{eqn:Gamma2}),
respectively. Hence in this case, if $|B-4|=O(1)$, the parametric resonance 
is efficient to produce the quanta of the corresponding Kaluza-Klein mode 
$\phi_{l_1l_2m_1m_2}$ only when 
%============< EQUATION >==============%
%
\begin{equation}
 \frac{\rho_{\sigma}}{\rho_0} \gtrsim 
	O(1)\cdot\left(\frac{\omega}{H}\right)^{3/2}.
\end{equation}
%======================================%

For $l_1$ and $l_2$ satisfying $B=4$, $N$ and $\Gamma$ are given by
(\ref{eqn:N2'}) and (\ref{eqn:Gamma2}) with $B=4$. 
Hence the parametric resonance is efficient to produce the quanta of
the corresponding Kaluza-Klein mode $\phi_{l_1l_2m_1m_2}$ when 
%============< EQUATION >==============%
%
\begin{equation}
 \frac{\rho_{\sigma}}{\rho_0} \gtrsim O(1)\cdot \frac{\omega}{H}.
\end{equation}
%======================================%

Finally for $l_1$ and $l_2$ satisfying $B=2$ or $B>4$, $N$ and $\Gamma$ 
are zero in our treatment. Thus, in this case, there is no regime when 
the the parametric resonance is efficient to produce the quanta of the 
corresponding Kaluza-Klein mode $\phi_{l_1l_2m_1m_2}$.

After all we can conclude that if there is a mode such that $B\le 1$ 
and $|B-1|=O(\epsilon)$ then the quanta of the Kaluza-Klein mode can 
be overproduced by the parametric resonance. Otherwise, at least in 
our situation ($H\ll\omega$ and $|\sigma /\sigma_{0}\ll1/D$) the parametric 
resonance is so mild that the overproduction does not occur.

%======================================%
%<<<<<<<<<<<< SECTION IV >>>>>>>>>>>>>>%
%======================================%

\section{Summary and discussion}\label{sec:discussion}

In this paper we have investigated a catastrophic creation of quanta 
of a Kaluza-Klein mode in a $D$-dimensional generalized Kaluza-Klein 
theory. As the origin of the catastrophic creation we have considered a 
small oscillation of a scale of compactification around a 
today's value of it. To make our arguments definite and for simplicity 
we have considered two classes of models of the compactification:
those by $S_{d}$ ($d=D-4$) and those by $S_{d_{1}}\times S_{d_{2}}$ 
($d_1\ge d_2$, $d_{1}+d_{2}=D-4$), then have shown that the 
compactification by $S_{d}$ can be stable for $d\ge 2$ and the 
compactification by $S_{d_{1}}\times S_{d_{2}}$ can be stable for 
$d_{1}\ge 2$. For these stable models we have given a 
hamiltonian for a Kaluza-Klein mode. A $4$-dimensional 
metric and a scale of the compactification have been treated 
classically, and the Kaluza-Klein mode quantum mechanically. The form of 
the hamiltonian for the Kaluza-Klein mode shows that a so-called parametric 
resonance phenomenon can occur: quanta of the Kaluza-Klein mode can be
excited catastrophically when frequency and amplitude of the 
oscillation of the scale of the compactification are in a band in their 
parameter space. We have given formulas of a creation rate and a 
number of created quanta of the Kaluza-Klein mode due to the 
parametric resonance, taking into account the first ($p/q=1$) and the
second ($p/q=2$) resonance band. After that by using the formulas we
have calculated  those quantities for each model of the
compactification.

We have shown that for the model of the compactification by
$S_{d}$ ($d\ge 2$) the first resonance cannot occur but the second
resonance can occur for the $l=1$ mode. 
For a  model of the compactification by $S_{d_{1}}\times S_{d_2}$, if
there exists a set of positive integers $l_1$ and $l_2$ satisfying
$B\le 1$ then the first resonance can also occur for the corresponding
mode $\phi_{l_1l_2m_1m_2}$, where $B$ is given by
(\ref{eqn:pareters-Sd1*Sd2}). Finally we have given conditions for the
parametric resonance to produce sufficient quanta of the Kaluza-Klein
mode for each model of the compactification. Our conclusion is as 
follows: (1) in the model of the compactification by $S_{d}$ the 
parametric resonance does not overproduce the quanta of the 
Kaluza-Klein mode; (2) in the model of the compactification by 
$S_{d_{1}}\times S_{d_{2}}$, if there is a mode satisfying $B\le 1$ 
and $|B-1|=O(\epsilon )$, then the quanta of the Kaluza-Klein mode 
can be overproduced by the parametric resonance; (3) otherwise in the 
$S_{d_{1}}\times S_{d_{2}}$ model, the parametric resonance is so mild 
that the overproduction does not occur.

As mentioned in Sec.~\ref{sec:intro}, if entropy production is 
negligible, there must not be a catastrophic creation of quanta of a 
Kaluza-Klein mode. This condition can be regarded as a cosmological 
criterion for compactification. Thus from the results obtained in this 
paper, assuming that entropy production is negligible, we can conclude 
that the model of the compactification by $S_{d_{1}}\times S_{d_{2}}$ is 
ruled out if there is a set of positive integers $l_{1}$ and $l_{2}$ 
satisfying the condition $B\le 1$ and $|B-1|=O(\epsilon )$, where $B$ is 
defined by (\ref{eqn:pareters-Sd1*Sd2}). One may expect that 
inflation followed by reheating can alter the conclusion and save the 
model. However, maybe it is not the case. In general it is natural 
that the field $\sigma$, which represents a scale of the 
compactification, couples to the inflaton field. Hence at the end of
the slow rolling phase of the inflation we can expect that energy
density $\rho_{\sigma}$ of coherent oscillation of the $\sigma$ is of
the same order as energy density of oscillation of the inflaton. In
this case the parametric resonance of a Kaluza-Klein mode occurs after
inflation, if our $4$-dimensional universe is obtained by the
compactification with the compact space $S_{d_{1}}\times S_{d_{2}}$
and if there is a set of positive integers $l_{1}$ and $l_{2}$
satisfying the condition $B\le 1$ and $|B-1|=O(\epsilon )$. 
The energy density $\rho_{KK}$ of the created quanta of the Kaluza-Klein 
mode is expected to be of the same order as $\rho_{\sigma}$. Hence 
$\rho_{KK}/\rho_{rad}=O(1)$ is expected after the reheating. Thus, naively, 
the inflation cannot save the model. However, we have to keep in mind that 
there are, as usual in arguments on early universe, loopholes in the above 
arguments: (1) a late time inflation may save the model if there is no
enough time after the late time inflation for the ratio 
$\rho_{KK}/\rho_{rad}$ to grow sufficiently; (2) if compactification 
radius is small compared with experimental scale but large enough compared 
with Planck scale then there may be a non-trivial range of allowed 
parameters. To fill up the loopholes is one of the future works we have 
to do.

Our next question now is whether or not the model of the compactification 
by $S_{d}$ and the model by $S_{d_{1}}\times S_{d_{2}}$, if there is 
no integers $l_{1}$ and $l_{2}$ satisfying $B\le 1$ and 
$|B-1|=O(\epsilon )$, do safely pass the criterion. We can say that the 
models are not ruled out from our analysis. However, unfortunately, we 
cannot answer this question definitely from only the results obtained in 
this paper since there is a possibility that the structure of the resonance 
may alter seriously when $\epsilon\gtrsim O(1)$. In this paper we have 
considered only the case that the scale of the compactification oscillates 
with a very small amplitude ($\epsilon\ll 1$). However, supposing that 
our $4$-dimensional universe was realized by a dynamical mechanism of 
compactification from a higher-dimensional spacetime, it is natural that 
there is an epoch when $\epsilon\gtrsim O(1)$ before the epoch we have 
considered. The former epoch is called broad resonance regime and in this 
regime it is expected that the resonance becomes broad and more efficient. 
In fact, in a simple model of reheating after inflation, it is the case 
\cite{Reheating}. Thus we have to analyze the former epoch in future works 
in order to answer the above question.

Although we have investigated only two classes of models of the
compactification, qualitative properties of parametric resonance of a
Kaluza-Klein mode are expected to be common in a wider class of models
of the compactification. It is because mass of a Kaluza-Klein mode
and frequency of an oscillation of a scale of the compactification are
both expected to be of order of the Planck mass in a wider class of
models. Hence it is valuable to analyze a parametric resonance
phenomenon of a Kaluza-Klein mode in more realistic models of the
compactification with other compact manifolds (eg. Calabi-Yau manifolds) 
or by other ways of stabilization of the compactification (eg. 
non-trivial configuration of anti-symmetric fields, gaugino condensation, 
etc.).

\vskip 1cm

\centerline{\bf Acknowledgments}
The author thanks Professor H. Kodama for continuous encouragement. He 
also thanks Professor J. Yokoyama for valuable comments.

%%%%%%%%%%%%%%%%%%%%%%%%%%%%%%%%%%%%%%%%%%%%%%%%%%%%%%%%%%%%%%%%%%%%
%%%%%%%%%%%%%%%%%%%%%%%%%%%%%%%%%%%%%%%%%%%%%%%%%%%%%%%%%%%%%%%%%%%%
% Appendix
%%%%%%%%%%%%%%%%%%%%%%%%%%%%%%%%%%%%%%%%%%%%%%%%%%%%%%%%%%%%%%%%%%%%
%%%%%%%%%%%%%%%%%%%%%%%%%%%%%%%%%%%%%%%%%%%%%%%%%%%%%%%%%%%%%%%%%%%%

\appendix

%======================================%
%<<<<<<<<<<<< APPENDIX A >>>>>>>>>>>>>>%
%======================================%

\section{Formula of the Floquet index}
\label{app:Floquet}

In this appendix we consider the following ordinary differential equation 
for a real function $Y$ of $z$.
%============< EQUATION >==============%
%
\begin{equation}
 \frac{d^2 Y}{dz^2} + h Y = g(z) Y, \label{eqn:Hill's-eq}
\end{equation}
%======================================%
where $h$ is a real constant and $g(z)$ is a real periodic function of 
$z$ with the period $2\pi$. We assume that $g(z)=g(-z)$. In this 
case this equation is called a Hill's equation. For a set of linearly 
independent two solutions $Y_{1}(z)$ and $Y_{2}(z)$, $Y_{1}(2\pi +z)$ 
and $Y_{2}(2\pi +z)$ are linearly independent, too. Hence, by the 
uniqueness of the solution of the equation, the later two solutions are 
linear combinations of the former two solutions: 
%============< EQUATION >==============%
%
\begin{equation}
 \left( \begin{array}{c}
	Y_{1}(2\pi +z) \\
	Y_{2}(2\pi +z)
 \end{array}	\right) = U
 \left( \begin{array}{c}
	Y_{1}(z) \\
	Y_{2}(z)
 \end{array}	\right),
\end{equation}
%======================================%
where $U$ is a $2\times 2$ regular matrix with all components being real 
constants. Therefore there are independent linear combinations 
$\tilde{Y}_{1}(z)$ and $\tilde{Y}_{2}(z)$ of $Y_{1}(z)$ and $Y_{2}(z)$ 
such that 
%============< EQUATION >==============%
%
\begin{eqnarray}
 \tilde{Y}_{1}(2\pi +z) & = & \phi_{1} \tilde{Y}_{1}(z),
 							\nonumber\\
 \tilde{Y}_{2}(2\pi +z) & = & \phi_{2} \tilde{Y}_{2}(z),
\end{eqnarray}
%======================================%
where $\phi_{1}$ and $\phi_{2}$ are eigenvalues of $U$ and are
not zero by definition. 
By introducing a constant $\mu$ by $\phi_{1}=e^{2\mu\pi}$, 
$\tilde{Y}_{1}(z)$ can be expressed by a periodic function $\phi(z)$ with 
the period $2\pi$ as ~\footnote{
Note that in general the function $\phi(z)$ is complex. When $\mu$ is real, 
$\phi(z)$ is a real function. 
}
%============< EQUATION >==============%
%
\begin{equation}
 \tilde{Y}_{1}(z) = e^{\mu z}\phi(z).
\end{equation}
%======================================%
Since (\ref{eqn:Hill's-eq}) is invariant under the replacement 
$z\to -z$, $\tilde{Y}_{1}(-z)$ is also a solution. Moreover, if $\Re\mu\ne 
0$ or $2\Im\mu$ is not an integer then $\tilde{Y}_{1}(-z)$ is linearly 
independent of $\tilde{Y}_{1}(z)$. Thus, in this case, 
$\phi_{2}=e^{-2\mu\pi}$ and the general solution of (\ref{eqn:Hill's-eq}) 
can be written as 
%============< EQUATION >==============%
%
\begin{equation}
 Y(z) = c_{1}e^{\mu z}\phi(z) + c_{2}e^{-\mu z}\phi(-z)
 	,\label{eqn:general-sol}
\end{equation}
%======================================%
where $c_{1}$ and $c_{2}$ are complex constants such that both $Y$ and 
$dY/dz$ are real at $z=z_{0}$~\footnote{
When $\mu$ is real, the coefficients $c_{1}$ and $c_{2}$ are arbitrary 
real constants.
}. 
The constant $\mu$ is called a Floquet index.

The equation (\ref{eqn:Hill's-eq}) is said to be stable when $\mu$ is
pure imaginary, while unstable when $\mu$ has a non-zero real
part~\cite{McLachlan}. 
The purpose of this appendix is to obtain a formula of the value of
$\mu$ and to give a criterion when the equation is unstable. For this
purpose we use the method of averaging~\cite{Bogoliubov}. Let us
suppose the case when the function $g(z)$ is parameterized by a small
parameter $\epsilon$ and $g(z)=O(\epsilon)$, and assume that $g(z)$
can be expanded by $\epsilon$. In this case by redefinition of the
real constant $h$ we can expand $g(z)$ as 
%============< EQUATION >==============%
%
\begin{equation}
 g(z) = \sum_{s=1}^{\infty}\sum_{n=-\infty}^{\infty}
	\epsilon^s g^{(s)}_{n}e^{inz}
	= 2\sum_{s=1}^{\infty}\sum_{n=1}^{\infty}
	\epsilon^s g^{(s)}_{n}\cos{nz}, 
 	\label{eqn:g-expand2}
\end{equation}
%======================================%
where $g^{(s)}_{n}$ $(s=1,2,\cdots; n=0,\pm 1,\pm 2,\cdots )$ are 
real constants satisfying
%============< EQUATION >==============%
%
\begin{eqnarray}
 g^{(s)}_{0} & = & 0,\nonumber\\
 g^{(s)}_{n} & = & g^{(s)}_{-n}
\end{eqnarray}
%======================================%
for all $s$.

Let us introduce a parameter $\Delta$ by
%============< EQUATION >==============%
%
\begin{equation}
 h = \left( \frac{p}{q}\right)^2 + \epsilon\Delta,
\end{equation}
%======================================%
where $p$ and $q$ are mutually prime positive integers. We consider 
a solution of the equation (\ref{eqn:Hill's-eq}) of the following 
form~\footnote{
Note that in order to obtain the formula of the value of $\mu$ we 
only have to seek a special solution of (\ref{eqn:Hill's-eq}) because 
of the form of the general solution (\ref{eqn:general-sol}). So we 
concentrate on this form of the solution. 
}.
%============< EQUATION >==============%
%
\begin{equation}
 Y(z) = a\cos\left(\frac{p}{q}z+\theta\right) 
 	+ \sum_{s=1}^{\infty}\epsilon^s u^{(s)}(a,\theta,z)
 	,\label{eqn:ansatz}
\end{equation}
%======================================%
where $a$ and $\theta$ are slowly varying real functions of $z$:
%============< EQUATION >==============%
%
\begin{eqnarray}
 \frac{da}{dz} & = & 
	\sum_{s=1}^{\infty}\epsilon^s A_{s}(a,\theta),\nonumber \\
 \frac{d\theta}{dz} & = & 
	\sum_{s=1}^{\infty}\epsilon^s B_{s}(a,\theta),
	\label{eqn:da-dtheta}
\end{eqnarray}
%======================================%
and $u^{(s)}$ $(s=1,2,\cdots)$ are functions satisfying
%============< EQUATION >==============%
%
\begin{equation}
 u^{(s)}(a,\theta,z) = u^{(s)}(a,\theta +2\pi,z) 
 	= u^{(s)}(a,\theta,z +2q\pi).
\end{equation}
%======================================%
Substituting the ansatz (\ref{eqn:ansatz}) into the Hill's equation
(\ref{eqn:Hill's-eq}), we obtain the following $O(\epsilon )$- and 
$O(\epsilon^2)$-equations.  
%============< EQUATION >==============%
%
\begin{eqnarray}
 \sum_{n=-\infty}^{\infty}
	\frac{p^2-n^2}{q^2}u_{n}^{(1)}(a,\theta)e^{inz/q} & = & 
	a\left(\sum_{n=-\infty}^{\infty}g_{n}^{(1)}e^{inz}
		+\frac{2p}{q}B_{1}(a,\theta)-\Delta\right)
		\cos\left(\frac{p}{q}z+\theta\right)\nonumber\\
 & & +\frac{2p}{q}A_{1}(a,\theta)\sin\left(\frac{p}{q}z+\theta\right) 
		,\label{eqn:O(epsilon)}  \\
 \sum_{n=-\infty}^{\infty}
	\frac{p^2-n^2}{q^2}u_{n}^{(2)}(a,\theta)e^{inz/q} & = & 
	\tilde{g}^{(2)}(a,\theta,z) \nonumber\\
 & & +\left(\frac{2p}{q}A_{2}+a\frac{\partial B_{1}}{\partial a}A_{1}
		+a\frac{\partial B_{1}}{\partial\theta}B_{1}+2A_{1}B_{1}
		\right)\sin\left(\frac{p}{q}z+\theta\right) \nonumber\\
 & & +\left(\frac{2p}{q}aB_{2}-\frac{\partial A_{1}}{\partial a}A_{1}
		-\frac{\partial A_{1}}{\partial\theta}B_{1}+aB_{1}^2
		\right)\cos\left(\frac{p}{q}z+\theta\right)
		,\label{eqn:O(epsilon2)} 
\end{eqnarray}
%======================================%
where the fourier components $u^{(1)}_{n}(a,\theta)$ and 
$u^{(2)}_{n}(a,\theta)$ of $u^{(1)}(a,\theta,z)$ and 
$u^{(2)}(a,\theta,z)$ have been introduced:
%============< EQUATION >==============%
%
\begin{equation}
 u^{(s)}(a,\theta,z) = 
	\sum_{n=-\infty}^{\infty}u^{(s)}_{n}(a,\theta)e^{inz/q},
\end{equation}
%======================================%
and $\tilde{g}^{(2)}$ is defined by
%============< EQUATION >==============%
%
\begin{eqnarray}
 \tilde{g}^{(2)}(a,\theta,z) & \equiv & 
	\left(\sum_{n=-\infty}^{\infty}g_n^{(1)}e^{inz}-\Delta\right)
	u^{(1)}(a,\theta,z)
 	+ a\sum_{n=-\infty}^{\infty}g_n^{(2)}e^{inz}
	\cos\left(\frac{p}{q}z+\theta\right)\nonumber\\
 & & - 2A_{1}(a,\theta)\frac{\partial^2u^{(1)}(a,\theta,z)}
 		{\partial a\partial z}
 	- 2B_{1}(a,\theta)\frac{\partial^2u^{(1)}(a,\theta,z)}
 		{\partial\theta\partial z}.
\end{eqnarray}
%======================================%

From the $O(\epsilon )$-equation (\ref{eqn:O(epsilon)}), we obtain the 
following form of $A_{1}$, $B_{1}$ and $u^{(1)}_{n}$ 
$(n\ne\pm p)$~\footnote{
When $n$ is not an integer we define $g^{(s)}_{n}$ by 
$g^{(s)}_{n}\equiv 0$  
}:
%============< EQUATION >==============%
%
\begin{eqnarray}
 A_{1}(a,\theta) & = & 
	-\frac{a}{2p/q}g^{(1)}_{2p/q}\sin{2\theta},\nonumber\\ 
 B_{1}(a,\theta) & = & 
 	\frac{1}{2p/q}\left[\Delta -g^{(1)}_{2p/q}\cos{2\theta}\right] 
		,\label{eqn:A1-B1}
\end{eqnarray}
%======================================%
and 
%============< EQUATION >==============%
%
\begin{equation}
 u^{(1)}_{n}(a,\theta) = \frac{aq^2}{2(p^2-n^2)}\left(
 	g^{(1)}_{n/q-p/q}e^{i\theta}+g^{(1)}_{n/q+p/q}e^{-i\theta}\right)
 	.
\end{equation}
%======================================%
Without loss of generality we can set $u^{(1)}_{\pm p}$ to be zero by 
redefinition of $a$ and $\theta$. Thus 
%============< EQUATION >==============%
%
\begin{equation}
 u^{(1)}(a,\theta,z) =
 	\sum_{n\ne\pm p}\frac{aq^2 g^{(1)}_{n/q+p/q}}{p^2-n^2}
 	\cos\left(\frac{n}{q}z-\theta\right).\label{eqn:u1}
\end{equation}
%======================================%

Next, from the $O(\epsilon^2)$-equation (\ref{eqn:O(epsilon2)}), we 
obtain 
%============< EQUATION >==============%
%
\begin{eqnarray}
 A_{2}(a,\theta) & = & 
	-\frac{1}{2p/q}\left( a\frac{\partial B_{1}}{\partial a}A_{1}
		+a\frac{\partial B_{1}}{\partial\theta}B_{1} + 2A_{1}B_{1}
		\right)
	-\frac{i}{2p/q}\left(\tilde{g}^{(2)}_{p}e^{-i\theta}
		- \tilde{g}^{(2)}_{-p}e^{i\theta}\right),\nonumber\\
 B_{2}(a,\theta) & = & 
 	\frac{1}{a\cdot 2p/q}\left(\frac{\partial A_{1}}{\partial a}A_{1}
		+\frac{\partial A_{1}}{\partial\theta}B_{1} - aB_{1}^2
		\right)
	-\frac{1}{a\cdot 2p/q}\left(\tilde{g}^{(2)}_{p}e^{-i\theta}
		+ \tilde{g}^{(2)}_{-p}e^{i\theta}\right)
		,\label{eqn:A2-B2}
\end{eqnarray}
%======================================%
where we have introduced the fourier component $\tilde{g}^{(2)}_{n}$ 
of $\tilde{g}^{(2)}$ by
%============< EQUATION >==============%
%
\begin{equation}
 \tilde{g}^{(2)}(a,\theta,z) = 
 	\sum_{n=-\infty}^{\infty}\tilde{g}^{(2)}_{n}(a,\theta)e^{inz/q}.
\end{equation}
%======================================%
Substituting (\ref{eqn:A1-B1}) and (\ref{eqn:u1}) into 
(\ref{eqn:A2-B2}), we obtain
%============< EQUATION >==============%
%
\begin{eqnarray}
 A_{2}(a,\theta) & = & 
	-\frac{a}{2p/q}\left[\sum_{n\ne 0,2p/q}
	\frac{g^{(1)}_{n} g^{(1)}_{2p/q-n}}{n(2p/q-n)}
	+g^{(2)}_{2p/q}\right]\sin{2\theta},\nonumber\\
 B_{2}(a,\theta) & = & 
	\frac{1}{(2p/q)^3}\left(g^{(1)2}_{2p/q}-\Delta^2\right)
	-\frac{1}{2p/q}\sum_{n\ne 0,2p/q}\frac{g^{(1)2}_{n}}{n(2p/q-n)}
	\nonumber\\
 & & -\frac{1}{2p/q}\left[\sum_{n\ne 0,2p/q}
	\frac{g^{(1)}_{n} g^{(1)}_{2p/q-n}}{n(2p/q-n)}
	+g^{(2)}_{2p/q}\right]\cos{2\theta}.\label{eqn:A2-B2'}
\end{eqnarray}
%======================================%

Finally we seek the formula of the value of $\mu$ up to order 
$O(\epsilon^2)$ by using the results obtained here. Let us introduce 
slowly varying functions $x$ and $y$ of $z$ by 
%============< EQUATION >==============%
%
\begin{eqnarray}
 x(z) & \equiv & a\cos\theta,\nonumber\\
 y(z) & \equiv & a\sin\theta.
\end{eqnarray}
%======================================%
The following equation for $x$ and $y$ can be obtained from 
(\ref{eqn:da-dtheta}), (\ref{eqn:A1-B1}) and (\ref{eqn:A2-B2'}). 
%============< EQUATION >==============%
%
\begin{equation}
 \frac{d}{dz}\left( \begin{array}{c}
	x(z) \\
	y(z)
 \end{array}	\right) = V \left( \begin{array}{c}
	x(z) \\
	y(z)
 \end{array} 	\right) + O(\epsilon^3),
\end{equation}
%======================================%
where $V$ is a $2\times 2$ matrix of the form 
%============< EQUATION >==============%
%
\begin{equation}
 V = \frac{\epsilon}{2p/q}\left( \begin{array}{cc}
	0 & -(\delta + \tilde{\Delta}) \\
	-(\delta -\tilde{\Delta}) & 0
 \end{array}	\right). \label{eqn:matrix-V}
\end{equation}
%======================================%
The constants $\delta$ and $\tilde{\Delta}$ are defined by
%============< EQUATION >==============%
%
\begin{eqnarray}
 \delta & = & g^{(1)}_{2p/q} +\epsilon\left[\sum_{n\ne 0,2p/q}
 	\frac{g^{(1)}_{n} g^{(1)}_{2p/q-n}}{n(2p/q-n)}
 	+g^{(2)}_{2p/q}\right] 	,\nonumber\\
 \tilde{\Delta} & = & \Delta + \epsilon\left[\frac{1}{(2p/q)^2}
 	\left(g^{(1)2}_{2p/q}-\Delta^2\right)
 	-\sum_{n\ne 0,2p/q}\frac{g^{(1)2}_{n}}{n(2p/q-n)}\right].
\end{eqnarray}
%======================================%
Since up to order $O(\epsilon^2)$ the Floquet index $\mu$ can be 
understood as an eigenvalue of $V$, it is given by
%============< EQUATION >==============%
%
\begin{equation}
 \mu = \mu_{\pm}
	= \pm \frac{|\epsilon|}{2p/q}\sqrt{\delta^2-\tilde{\Delta}^2}
 	+O(\epsilon^3)	.\label{eqn:mu}
\end{equation}
%======================================%
Thus the necessary and sufficient condition for the equation 
(\ref{eqn:Hill's-eq}) to be unstable is  
%============< EQUATION >==============%
%
\begin{equation}
 -|\delta | < \tilde{\Delta} < |\delta |.
\end{equation}
%======================================%
This inequality can be solved with respect to $\Delta$ by using the 
fact that $\Delta =0(1)$. The result is as follows up to 
$O(\epsilon)$. 
%============< EQUATION >==============%
%
\begin{equation}
 \Delta_{0}-|\delta| < \Delta < \Delta_{0}+|\delta|	,
\end{equation}
%======================================%
where $\Delta_{0}$ is defined by 
%============< EQUATION >==============%
%
\begin{equation}
 \Delta_{0} = 
 	\epsilon\sum_{n\ne 0,2p/q}\frac{g^{(1)2}_{n}}{n(2p/q-n)}.
\end{equation}
%======================================%

%======================================%
%<<<<<<<<<<<< APPENDIX B >>>>>>>>>>>>>>%
%======================================%
\section{Number of quanta excited by parametric resonance}
\label{app:number-quanta}

In this appendix we derive a formula of number of quanta excited by the 
parametric resonance. Although the formula can be found in 
literature~\footnote{
For example, see Ref. \cite{STB}. However, their derivation is
incomplete as will be stated later. The obtained formula of the number 
of quanta is exactly same in Ref. \cite{STB} and in this appendix.
}, 
we derive it in this appendix for completeness. 

We consider a quantum mechanics of an oscillator described by the time 
dependent Hamiltonian 
%============< EQUATION >==============%
%
\begin{equation}
 \hat{H} = \frac{1}{2}\left[\hat{P}^2+\Omega^2(t)\hat{Q}^2\right], 
\end{equation}
%======================================%
where $\Omega(t)$ is a positive function of time $t$, $\hat{Q}$ and 
$\hat{P}$ are operators representing a coordinate of the oscillator and 
its conjugate momentum: 
%============< EQUATION >==============%
%
\begin{equation}
 \left[\hat{Q},\hat{P}\right] = i.
\end{equation}
%======================================%

First let us define the time dependent operator $\hat{a}(t)$ by 
%============< EQUATION >==============%
%
\begin{equation}
 \hat{a}(t) = \frac{e^{i\int\Omega dt}}{\sqrt{2\Omega}}
 	\left(\Omega\hat{Q}+i\hat{P}\right). \label{eqn:def-a}
\end{equation}
%======================================%
The operator $\hat{a}(t)$ and its hermitian conjugate 
$\hat{a}^{\dagger}(t)$ have the standard commutation relation of 
creation and annihilation operators:
%============< EQUATION >==============%
%
\begin{equation}
 \left[\hat{a},\hat{a}^{\dagger}\right] = 1.
\end{equation}
%======================================%
Since the hamiltonian can be expressed as
%============< EQUATION >==============%
%
\begin{equation}
 \hat{H} = \Omega\left(\hat{a}^{\dagger}\hat{a}+\frac{1}{2}\right)
 	,
\end{equation}
%======================================%
the time dependent vacuum state $|0_{t}\rangle$, which is defined by 
$\hat{a}|0_{t}\rangle =0$, minimizes the value of the Hamiltonian at a
moment $t$. Hence, the creation and the annihilation operators
$\hat{a}(t)$ and $\hat{a}^{\dagger}(t)$ regard, so to speak, 
instantaneous quanta at the moment $t$.

The equation of motion for $\hat{Q}$ and $\hat{P}$ is equivalent to 
the following differential equation for the operator
$\hat{a}(t)$~\footnote{
Hereafter the dot denotes the derivation with respect to the time
coordinate $t$. 
}:
%============< EQUATION >==============%
%
\begin{equation}
 \dot{\hat{a}} = \frac{\dot{\Omega}}{2\Omega}e^{2i\int\Omega dt}
 	\hat{a}^{\dagger} .
\end{equation}
%======================================%
A general solution of this equation is written as
%============< EQUATION >==============%
%
\begin{equation}
 \hat{a}(t) = \alpha(t)\hat{a}(0) + 
	\beta^*(t)\hat{a}^{\dagger}(0),\label{eqn:a-alpha-beta}
\end{equation}
%======================================%
where $\alpha(t)$ and $\beta(t)$ are complex functions satisfying the 
differential equation
%============< EQUATION >==============%
%
\begin{eqnarray}
 \dot{\alpha} & = & 
 	\frac{\dot{\Omega}}{2\Omega}e^{2i\int\Omega dt}\beta
 	,\nonumber\\
 \dot{\beta} & = & 
 	\frac{\dot{\Omega}}{2\Omega}e^{-2i\int\Omega dt}\alpha
 	,
\end{eqnarray}
%======================================%
and the initial condition
%============< EQUATION >==============%
%
\begin{equation}
 \alpha(0) =1,\ \ \beta(0) =0.\label{eqn:init-alpha-beta}
\end{equation}
%======================================%
The number of the instantaneous quanta at the moment $t$ for the vacuum 
state $|0_{0}\rangle$, which is annihilated by $\hat{a}(0)$, can be  
expressed by $\beta(t)$ as
%============< EQUATION >==============%
%
\begin{equation}
 N(t) = \langle 0_{0}|\hat{a}^{\dagger}(t)\hat{a}(t)|0_{0}\rangle 
 	= \left|\beta(t)\right|^2.\label{eqn:N-beta}
\end{equation}
%======================================%

By using (\ref{eqn:def-a}) and (\ref{eqn:a-alpha-beta}), the 
operator $\hat{Q}$ is expressed by the operator $\hat{a}(0)$ and its 
hermitian conjugate $\hat{a}^{\dagger}(0)$ as
%============< EQUATION >==============%
%
\begin{equation}
 \hat{Q}(t) = Q^{(+)}(t)\hat{a}(0) + 
	 Q^{(-)}(t)\hat{a}^{\dagger}(0),  
\end{equation}
%======================================%
where $Q^{(+)}(t)$ and $Q^{(-)}(t)$ are defined by
%============< EQUATION >==============%
%
\begin{eqnarray}
 Q^{(+)}(t) & = & 
 	\frac{1}{\sqrt{2\Omega(t)}}\left[
 	\alpha(t)e^{-i\int\Omega dt}+\beta(t)e^{i\int\Omega dt}\right]
 	,\nonumber\\
 Q^{(-)}(t) & = & 
 	\frac{1}{\sqrt{2\Omega(t)}}\left[
 	\beta^*(t)e^{-i\int\Omega dt}+\alpha^*(t)e^{i\int\Omega dt}
	\right]	,
\end{eqnarray}
%======================================%
or equivalently,
%============< EQUATION >==============%
%
\begin{eqnarray}
 \alpha(t) = \frac{e^{i\int\Omega dt}}{\sqrt{2\Omega(t)}}
 	\left[\Omega(t)Q^{(+)}(t)+i\dot{Q}^{(+)}(t)\right]
	,\nonumber\\
 \beta(t) = \frac{e^{-i\int\Omega dt}}{\sqrt{2\Omega(t)}}
 	\left[\Omega(t)Q^{(-)}(t)-i\dot{Q}^{(-)}(t)\right].
 	\label{eqn:alpha-beta-Q}
\end{eqnarray}
%======================================%
Since $\hat{a}(0)$ and $\hat{a}^{\dagger}(0)$ do not commute with each 
other, the equation of motion of $\hat{Q}(t)$ means
%============< EQUATION >==============%
%
\begin{equation}
 \ddot{Q}^{(\pm)} + \Omega^2 Q^{(\pm)} = 0.\label{eqn:eq-Qpm}
\end{equation}
%======================================%
It is shown that the initial condition (\ref{eqn:init-alpha-beta}) is 
equivalent to  
%============< EQUATION >==============%
%
\begin{equation}
 Q^{(-)}(0) = \frac{1}{\sqrt{2\Omega(0)}},\ \ 
 \Omega(0)Q^{(-)}(0)-i\dot{Q}^{(-)}(0) = 0 ,\label{eqn:init-Q}
\end{equation}
%======================================%
by using the relation 
%============< EQUATION >==============%
%
\begin{equation}
 \alpha (t) = \frac{2\Omega(t)}{\dot{\Omega}(t)}
 	e^{2i\int\Omega dt}\dot{\beta}(t)
	= \frac{e^{i\int\Omega dt}}{\sqrt{2\Omega(t)}}\left(
 	\Omega(t)Q^{(-)}(t) + i\dot{Q}^{(-)}(t)\right).
\end{equation}
%======================================%
Thus, in order to calculate the number of instantaneous quanta $N(t)$,
what we have to do is to seek the solution $Q^{(-)}(t)$ of the
differential equation (\ref{eqn:eq-Qpm}) with the initial condition
(\ref{eqn:init-Q}), and to substitute it into (\ref{eqn:N-beta}) with 
(\ref{eqn:alpha-beta-Q}). Note that the function $Q^{(-)}$ is complex
while the classical coordinate $Q$ corresponding to the operator
$\hat{Q}$ is real.

Let us consider the case when the function $\Omega(t)$ is given by
%============< EQUATION >==============%
%
\begin{equation}
 \Omega^2(t) = \omega^2[h-g(\omega t)],
\end{equation}
%======================================%
where $\omega$ and $h$ are positive constants, $g(z)$ is a real
function given by (\ref{eqn:g-expand2}). In this case the corresponding 
differential equation for $Q^{(-)}$ is the Hill's equation
(\ref{eqn:Hill's-eq}) with $Y$ replaced by $Q^{(-)}$, provided that 
%============< EQUATION >==============%
%
\begin{equation}
 z=\omega t. 
\end{equation}
%======================================%
In Appendix \ref{app:Floquet} we have obtained a real solution of the 
form $(\ref{eqn:ansatz})$. We write the real part and the imaginary 
part of $Q^{(-)}$ in the form $(\ref{eqn:ansatz})$, respectively:
%============< EQUATION >==============%
%
\begin{equation}
 Q^{(-)}(t) = (x_{1}+ix_{2})\cos\left(\frac{p}{q}z\right)
 	- (y_{1}+iy_{2})\sin\left(\frac{p}{q}z\right) 
 	+ O(\epsilon) 	,\label{eqn:Q-x-y}
\end{equation}
%======================================%
where $x_{1,2}$ and $y_{1,2}$ are real functions satisfying the 
following differential equations.
%============< EQUATION >==============%
%
\begin{equation}
 \frac{d}{dz}\left( \begin{array}{c}
	x_{1,2}(z) \\
	y_{1,2}(z)
 \end{array}	\right) = V \left( \begin{array}{c}
	x_{1,2}(z) \\
	y_{1,2}(z)
 \end{array} 	\right) + O(\epsilon^3),
	\label{eqn:eq-x12-y12}
\end{equation}
%======================================%
where the matrix $V$ is given by (\ref{eqn:matrix-V}). For the
functions $x_{1,2}$ and $y_{1,2}$, the following initial condition in
the lowest order of $\epsilon$ is derived from the initial condition
(\ref{eqn:init-Q}):
%============< EQUATION >==============%
%
\begin{eqnarray}
 x_{1}(0) & = & y_{2}(0) = \frac{1}{\sqrt{\omega\cdot 2p/q}}
 	,\nonumber\\
 x_{2}(0) & = & y_{1}(0) = 0.
\end{eqnarray}
%======================================%
Hence, in the lowest order, the solution of (\ref{eqn:eq-x12-y12}) is
given by 
%============< EQUATION >==============%
%
\begin{eqnarray}
 x_{1} & = & \frac{1}{\sqrt{\omega\cdot 2p/q}}
 	\cosh{\mu_+z},\nonumber\\
 y_{1} & = & -\frac{1}{\sqrt{\omega\cdot 2p/q}}
 	\sqrt{\frac{\delta +\tilde{\Delta}}
 	{\delta -\tilde{\Delta}}}\sinh{\mu_+z}
 	,\nonumber\\
 x_{2} & = & -\frac{1}{\sqrt{\omega\cdot 2p/q}}
	\sqrt{\frac{\delta -\tilde{\Delta}}
	{\delta +\tilde{\Delta}}}\sinh{\mu_+z}
 	,\nonumber\\
 y_{2} & = & \frac{1}{\sqrt{\omega\cdot 2p/q}}
 	\cosh{\mu_+z}.\label{eqn:sol-x-y}
\end{eqnarray}
%======================================%
Since it is shown from (\ref{eqn:alpha-beta-Q}) and (\ref{eqn:Q-x-y})
that 
%============< EQUATION >==============%
%
\begin{equation}
 |\beta(t)| = \frac{\sqrt{\omega\cdot 2p/q}}{2}
 	\sqrt{(x_{1}-y_{2})^2+(x_{2}+y_{1})^2}+0(\epsilon),
\end{equation}
%======================================%
the number of quanta $N(t)$ defined by (\ref{eqn:N-beta})
is~\footnote{
In Ref. \cite{STB} they used the equation
$\ddot{\beta}=\mu_+^2\beta+O(\epsilon)$ to derive this formula of
$N(t)$. However, the term $\mu_+^2\beta$ is of order
$O(\epsilon^2)$ since $\mu_{+}$ is of order $O(\epsilon)$. Thus their 
derivation is incomplete. 
}
%============< EQUATION >==============%
%
\begin{equation}
 N(t) = \frac{1}{1-\left(\tilde{\Delta}/\delta\right)^2}
 	\sinh^2{\mu_+z}+0(\epsilon).
\end{equation}
%======================================%

Finally we consider the case when the value of $\delta$ and 
$\tilde{\Delta}$ are not constants in time but they are given as
functions of $z$. We assume that their change is so slow that 
%============< EQUATION >==============%
%
\begin{equation}
 \left|\frac{d}{dz}\ln |\delta\pm\tilde{\Delta}|\right|
	\ll \mu_+.
\end{equation}
%======================================%
In this case $\mu_+z$ in (\ref{eqn:sol-x-y}) is replaced by 
$\int\mu_+dz$. Thus the number of quanta in this case is
%============< EQUATION >==============%
%
\begin{equation}
 N(t) \simeq \sinh^2\left(\int\mu_+dz\right).
\end{equation}
%======================================%

%======================================%
%<<<<<<<<<<<< REFERENCES >>>>>>>>>>>>>>%
%======================================%


\begin{thebibliography}{22}
\bibitem{GSW}
M. B. Green, J. H. Schwartz, E. Witten, {\it Superstring theory}
(Cambridge University Press, 1987).
\bibitem{M-theory}
J. H. Schwarz, 'Lectures on Superstring and M Theory Dualities',
hep-th/9607201, and references therein. 
\bibitem{Kolb&Slansky}
E. W. Kolb and R. Slansky, Phys. Lett. {\bf B135}, 378 (1984).
\bibitem{Landau}
L. Landau and E. Lifschitz, {\it Mechanics} (Pergamon, Oxford, 1960).
\bibitem{Reheating}
For example, L. Kofman, A. Linde and A. A. Starobinsky, Phys. Rev. 
{\bf D56}, 3258 (1997), and references therein.
\bibitem{STB}
Y. Shtanov, J. Traschen and R. Brandenberger, Phys. Rev. {\bf D51}, 
5438 (1995). 
\bibitem{Candelas&Weinberg}
P. Candelas and S. Weinberg, Nucl. Phys. {\bf B237}, 397 (1984).
\bibitem{Potential-U1}
For example, K. Maeda, Phys. Lett. {\bf B186}, 33 (1987); L. Amendola, 
E. W. Kolb, M. Litterio and F. Occhionero, Phys. Rev. {\bf D42}, 1944 
(1990). 
\bibitem{McLachlan}
N. W. McLachlan, {\it Theory and Application of Mathieu Functions} 
(Oxford University Press, London, 1947). 
\bibitem{Bogoliubov}
N. Bogoliubov and Y. Mitropolsky, {\it Asymptotic Methods in the 
Theory of NonLinear Oscillations} (Hindustan Publishing, Delhi, 1961). 
\end{thebibliography}
\end{document}